\documentclass[sigconf, 10pt]{acmart}

\usepackage{mathtools}
\usepackage{wrapfig}
\usepackage[ruled,vlined]{algorithm2e}
\usepackage{colortbl}
\usepackage{url}
\usepackage{color,soul}
\usepackage{graphics}
\usepackage{breakurl}
\usepackage{lipsum}
\usepackage{tikz}
\usepackage{listings} 
\usepackage{euler}

\usepackage{booktabs}	
\usepackage{lipsum}
\usepackage{subcaption}
\usepackage{boldline}
\usepackage{pifont} 
\usepackage{capt-of}	
\usepackage{todonotes}

\usepackage{adjustbox}

\usepackage{setspace}

\definecolor{labelkey}{rgb}{0,1,0}

\definecolor{airforceblue}{rgb}{0.36, 0.54, 0.66}
\definecolor{frenzyorange}{RGB}{249, 158, 26}

\renewcommand{\paragraph}[1]{\vskip 3pt\noindent\textbf{#1 }}	 
  {\begin{list}{$\bullet$}%
     {\setlength{\parsep}{0pt}%
      \setlength{\topsep}{0pt}%
      \setlength{\itemsep}{2pt}}}%
  {\end{list}}
\newcommand\note[1]{\sethlcolor{yellow} \hl{#1}} 
\newcommand\Note[1]{\sethlcolor{green} \hl{#1}} 
\newcommand\noted[1]{} 

  {\begin{itemize}
	[leftmargin=0cm,itemindent=.3cm,labelwidth=\itemindent,
		labelsep=0pt,
		parsep=0pt,
		topsep=1pt,
		itemsep=2pt,
		align=left]
  }%
  {\end{itemize}}    

  {\begin{enumerate}
	[leftmargin=0cm,itemindent=.5cm,labelwidth=\itemindent,
		labelsep=0pt,
		parsep=1pt,
		topsep=1pt,
		itemsep=3pt,
		align=left]
  }%
  {\end{enumerate}}

  {\begin{enumerate*}
	[label=\roman*)]
  }%
  {\end{enumerate*}}

\newcommand\sect[1]{Section~\ref{sec:#1}}	
        
\newcommand{\sys}[1]{SwapNN}

\renewcommand\footnotetextcopyrightpermission[1]{} 
\usepackage{booktabs} 

\begin{document}
 
\title{Enabling Large Neural Networks on Tiny Microcontrollers with Swapping}

\author{Hongyu Miao}
\affiliation{%
	\institution{Purdue ECE}
}

\author{Felix Xiaozhu Lin}
\affiliation{%
	\institution{University of Virginia}
}

\begin{abstract}
Running neural networks (NNs) on microcontroller units (MCUs) is becoming increasingly important, but is very difficult due to the tiny SRAM size of MCU. Prior work proposes many algorithm-level techniques to reduce NN memory footprints, but all at the cost of sacrificing accuracy and generality, which disqualifies MCUs for many important use cases. We investigate a system solution for MCUs to execute NNs out of core: dynamically swapping NN data chunks between an MCU's tiny SRAM and its large, low-cost external flash. Out-of-core NNs on MCUs raise multiple concerns: execution slowdown, storage wear out, energy consumption, and data security. We present a study showing that none is a showstopper; the key benefit -- MCUs being able to run large NNs with full accuracy and generality -- triumphs the overheads. Our findings suggest that MCUs can play a much greater role in edge intelligence.

\end{abstract}



\maketitle

\section{Introduction}


With low cost and energy, MCUs are becoming ubiquitous platforms for neural networks (NNs), a paradigm dubbed tinyML~\cite{introtinyml}. 
Running NN \textit{on MCU}, rather than sending raw data off, offers multiple advantages, notably tolerating poor networks and preserving data privacy. 
Use cases include 
detecting farming crop disease by classifying leaf photos~\cite{nuruai} and
extracting traffic patterns by analyzing city images. 




\begin{figure}[t!]
\centering
    \includegraphics[width=0.45\textwidth{}]{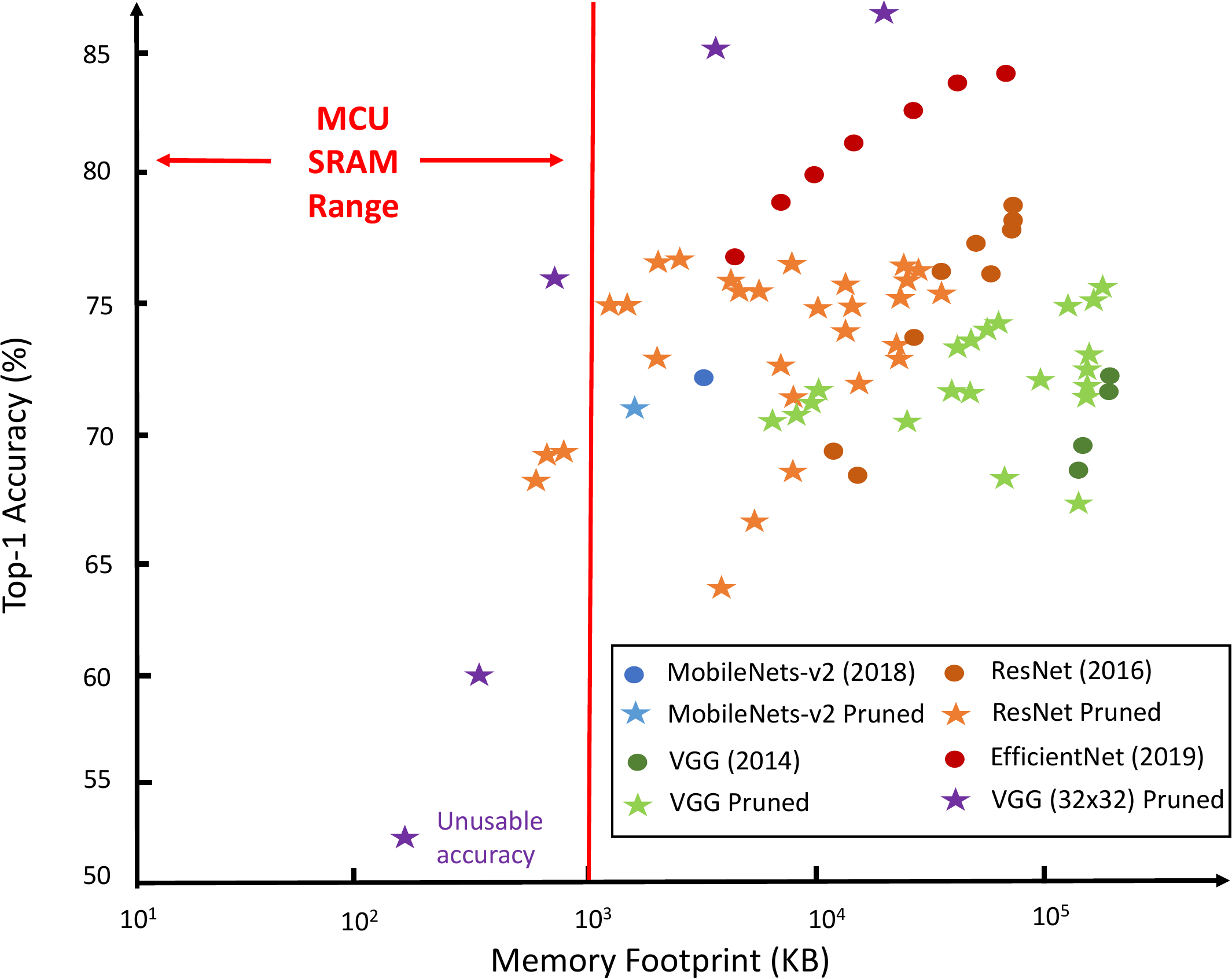} 
    \caption{Many popular NNs exceed the MCU memory size~\cite{statofnnpruning}.}
    \label{fig:models-accuracy-memory}
\end{figure}

A top obstacle in tinyML is memory limit. 
On one hand, an MCU has small memory, which comprises
tens to hundreds KB of SRAM as the main memory and 
byte-addressable flash of no more than a few MBs for read-only data. 
Note that the byte-addressable flash is different from external block-addressable storage such as SD cards~\cite{stm32}. 

On the other hand, 
state-of-the-art NNs achieve high accuracy and generality with large memory footprints~\cite{googlenet, siu2018memory}. 
An NN's memory footprint includes read-only parameters and intermediate/final results called feature maps. 
Although MCU can process one NN layer in memory before loading the next layer, 
a layer's parameters and feature maps can still take up to 100 MB (e.g. VGG16~\cite{vgg}). 
This exceeds the MCU memory size by up to two orders of magnitude. 
Such a memory gap is widening as recent NNs are becoming larger~\cite{szegedy2015going} while MCU memory sees slow, if at all, scaling due to cost constraints~\cite{sramiot}. 

A popular approach to overcoming memory limitation is to engineer NNs themselves. 
Common techniques include model compression~\cite{prune, deepcompression, prunefilter}, parameter quantization~\cite{limitedprecision}, designing tiny NNs from scratch~\cite{bmxnet}, as well as automation of these procedures~\cite{lin2020mcunet}. 
In exchange, this approach gives away model accuracy or generality at varying degrees. 
Unfortunately, in order for an NN to fit into the MCU memory, 
the NN either becomes substantially inaccurate (e.g. $<$ 60\% top-1 accuracy as shown in Figure~\ref{fig:models-accuracy-memory}) or too specialized (e.g. can only detect a few object classes~\cite{xu2020approximate}).

This disqualifies MCUs from the use cases where high accuracy/generality are desired while delays can be tolerated, for example: 
(1) \textit{NN inference on slowly changing signals}, 
e.g., monitoring crop health by analyzing hourly photos~\cite{nuruai} and traffic patterns by analyzing video frames every 20-30 minutes~\cite{xu2020approximate}.
(2) \textit{profiling NNs on device}: occasionally running a full-blown NN to estimate the accuracy of long-running smaller NNs~\cite{shen2017fast}; 
(3) \textit{transfer learning}: re-training NNs on MCUs with data collected from deployment every hour or day~\cite{deeptype}. 


\paragraph{A case for out-of-core NNs}
Can an MCU execute NNs that far exceed its physical memory size? 
A proven wisdom is to dynamically swap \textit{tiles} of NN layers between memory tiers~\cite{layerfusion}. 
Specially, an MCU runtime can split one NN layer's working set into a series of tiles, each small enough to fit the MCU memory; load tiles from external storage (a micro SD card) to memory, compute on them, and write results back to the storage for subsequent processing. 
While prior systems have swapped NN tiles between a server's CPU/GPU memories ~\cite{swapadvisor}, 
applying the idea to MCU, in particular swapping between small SRAM and a wimpy SD card, 
raises multiple concerns: loss of SD card durability, execution slowdown due to IO operations, energy increase, and safety/security of out-of-core NN data.
This paper aims to address these concerns.



\paragraph{Key observations}
This paper demonstrates the practicality of out-of-core NN on MCUs, 
for which we have following observations. 


\begin{itemize}
	\item 
	\textit{Swapping overhead is only pronounced in certain NN layers.} \hspace{2mm}
	Only on layers with low arithmetic intensity, notably fully connected (FC) layers, the swapping delay due to IO is longer than that of computation; 
	on layers with higher arithmetic intensity, e.g. convolution (Conv), 
	the swapping delay is dwarfed by that of computation.
	The swapping overhead is further diminished by MCU's relative low CPU speed as compared to its IO speed.

	\item 
	\textit{Swapping rate is throttled by computation}, which limits the wear rate of SD cards.
	As a common NN structure, IO-bound layers such as FC are spaced by compute-bound layers such as Conv. 
	As a result, even with continuous NN executions, IO is only exercised intermittently. 

	\item 
	\textit{Most IO traffic for swapping is read}  \hspace{2mm}
	This is because a layer's parameters and input feature maps are often much larger than its output feature maps. 
	Fortunately, read traffic does not wear SD cards. 
	

	
	\item \textit{Hide swapping delays with parallelism at various granularities}. \hspace{2mm}
	Within a layer, the MCU can exploit \textit{tile} parallelism, by computing on a tile while transferring others to/from the storage.
	Between consecutive NN executions such as on a sequence of video frames, the MCU can further exploit \textit{pipeline} parallelism, by overlapping the swapping IO for an earlier frame with the computation of a later frame.

	\item 
	\textit{Modern MCU hardware often over-provision durability}. \hspace{2mm} 
	For example, a 64 GB SD card can last more than 10 years with 100 GB of daily writes (\sect{durability}). 
	As such, MCU can trades the surplus durability as a system resource for accommodating large NNs. 
	Modern MCUs incorporate rich specialized hardware, e.g., for DMA, hash, and crypto, which accelerates and secures IO operations. 
	
	\item 
	\textit{IO adds marginal energy to an already busy MCU.} \hspace{2mm}
	With an MCU already busy on computation, most of its hardware components in high power states. Further activating the SD card increases the system energy moderately. 

\end{itemize}

\paragraph{Quantitative findings}
We present \sys{}, a scheduler design that automatically schedules IO and compute tasks.
\sys{} exploits the IO/compute parallelism across tiles, layers, and data frames, meanwhile respects memory constraint and data dependency.
We applied \sys{} to a diverse set of NNs, MobileNets~\cite{mobilenets}, AlexNet~\cite{alexnet}, and VGG16~\cite{vgg}, on a Cortex-M7 MCU with 340 KB of SRAM. Our findings are: 

\begin{itemize}
	\item \textit{Low to modest speed overhead.} \hspace{2mm} NNs with dominant compute-bound layers see negligible swapping overhead, both in per-frame delay and frame throughput. 
	Compared to running VGG on an \textit{ideal} MCU with infinite main memory (SRAM), out-of-core execution with 512 KB memory sees only 6.9\% longer per-frame delay and only 3\% lower throughput.
	NNs with more IO-bound layers such as AlexNet see notable delay increase  (50\%) while insignificant loss in throughput (15.7\%) thanks to tile and pipeline parallelism. 
	
	\item \textit{Large tiles are crucial to low swapping overhead.}
	A key parameter in out-of-core NN is the tile size, which determines the granularity of IO/compute task. 	
	While small tiles lead to fine-grained tasks and therefore better  compute/IO parallelism, 
	they increase the total amount of IO traffic and the per-byte IO delay. 
	As we will show experimentally, the cost of small tiles overshadows the benefit of parallelism on typical MCU hardware and NNs, 

	\item \textit{Low durability loss.} \hspace{2mm}
	Even with an MCU executing NNs continuously, the write traffic due to swapping is no more than a few hundred GBs per day, comparable to SD card writes on a commodity surveillance camera.   
	%
	A 64 GB SD card can sustain such a write rate for 7.5 years before half of its cells are worn out. 
	
	\item \textit{Modest increase in energy consumption.} \hspace{2mm}
	Our \textit{worst-case} estimation shows 
	swapping increases system energy by less than 42\% compared to running NNs with infinite memory (all in memory without swapping). 
	
	\item \textit{Out-of-core data can be secured} with known mechanisms, such as encryption and hash-based integrity protection.
	Specialized hardware on MCUs further reduces their overhead. 
\end{itemize}


\paragraph{Contributions} 
Our contributions are as follows. 
\begin{itemize}
	\item We present the first study of applying swapping to NN on MCUs.
	We analyze the swapping-generated IO activities and their implications on performance, storage durability, energy, and data security.


	
	\item
	We explore software/hardware parameters that impact swapping overhead. 
	Towards lowering swapping overhead, 
	our findings shed light on setting software parameters and designing MCU hardware (e.g., choosing SRAM size).

	\item
	We present a scheduler design that can automatically schedule IO and compute tasks in parallel. 
	The scheduler exploits a common NN characteristic that an NN often has a mix of IO-bound and compute-bound layers.
	It exploits IO/compute parallelism across NN layers and across data frames while respecting memory constraint and data dependency. 

	 \item 
	 We make a case that an MCU of less than ten dollars with 
	 hundreds of KB SRAM
	 can execute large NNs such as VGG16, which expands the scope of tinyML significantly. 
	 
\end{itemize}

\section{Background and Motivations}
\subsection{A taxonomy of NN layers}
\label{sec:io-compute}

\begin{table}[t!]
\centering
    \includegraphics[width=0.40\textwidth{}]{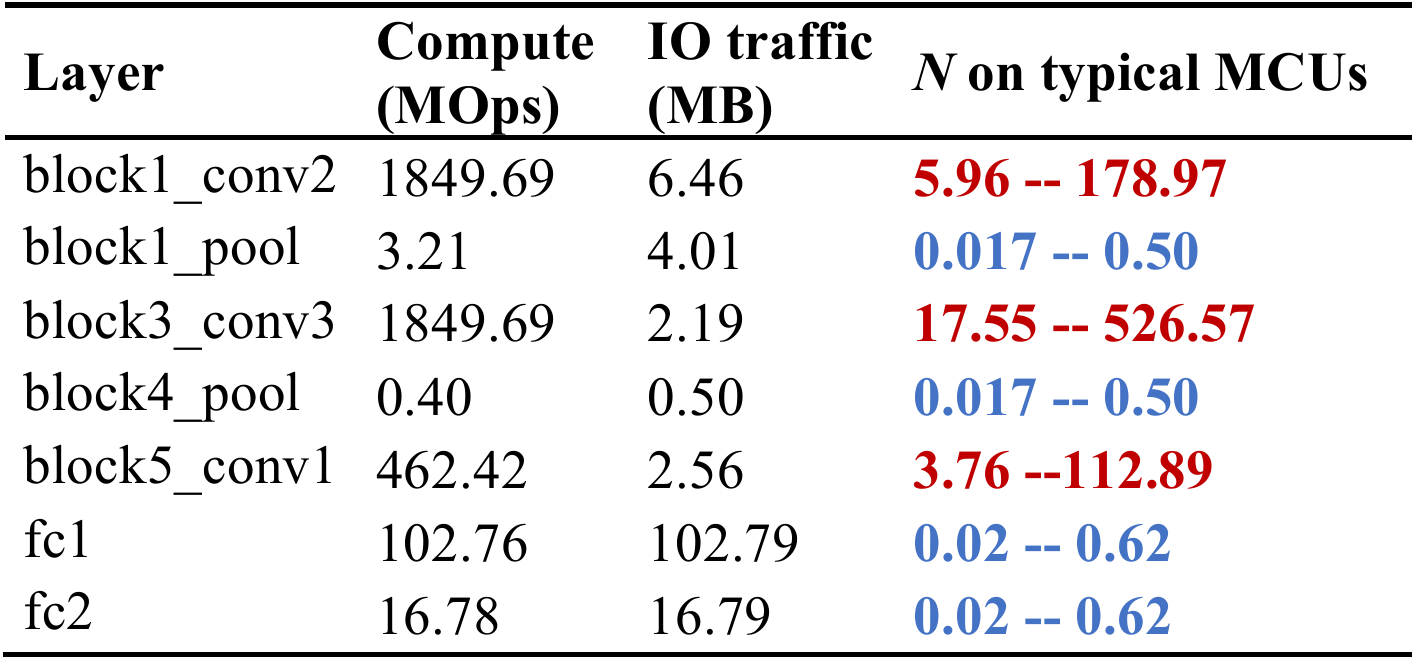} 
    \caption{Normalized arithmetic intensity ($N$) on NN layers with MCU's common speed range (64--480 MOPS~\cite{flops,cortexm}) and IO bandwidth range (10--40 MB/s~\cite{microsdbw}). NN: VGG16}
    \label{tab:nn-layers-arithmetic-intensity}
\end{table}

\begin{table}[t!]
\centering
    \includegraphics[width=0.48\textwidth{}]{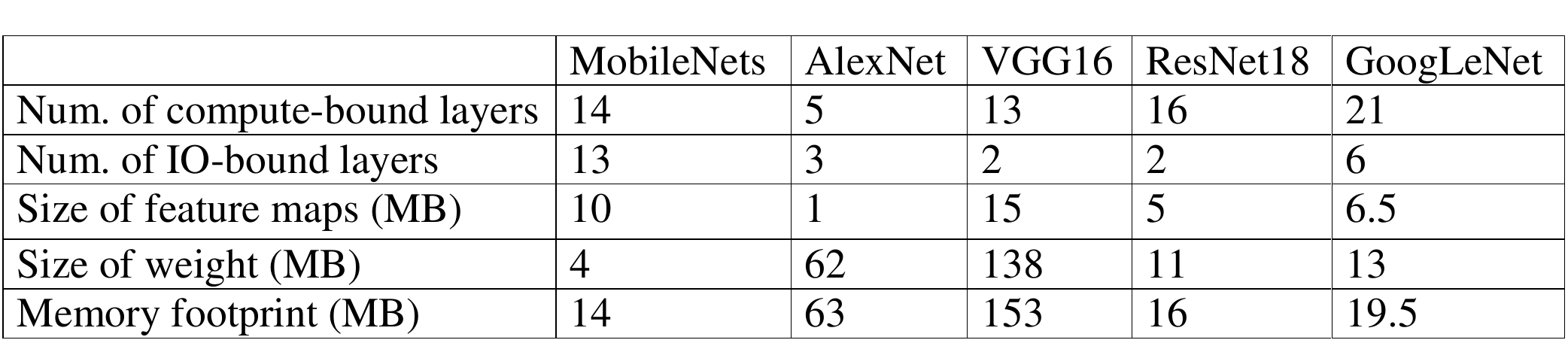} 
    \caption{Number of IO-bound and compute-bound layers and quantized memory footprints of popular NNs~\cite{convnet-burden}.} 
    \label{tab:cnn-study}
\end{table}

\begin{figure*}
    \centering
    \begin{subfigure}[b]{0.28\textwidth}
        \includegraphics[width=\textwidth]{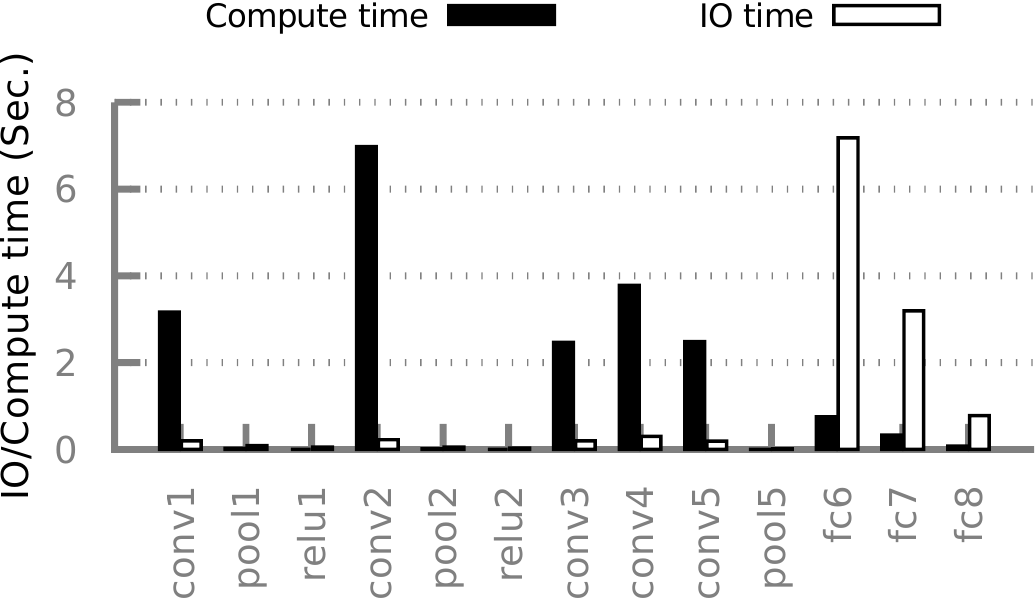}
        \caption{AlexNet (input shape: 227)}
        \label{fig:alexnet-128k}
    \end{subfigure}
    ~ 
    \begin{subfigure}[b]{0.33\textwidth}
        \includegraphics[width=\textwidth]{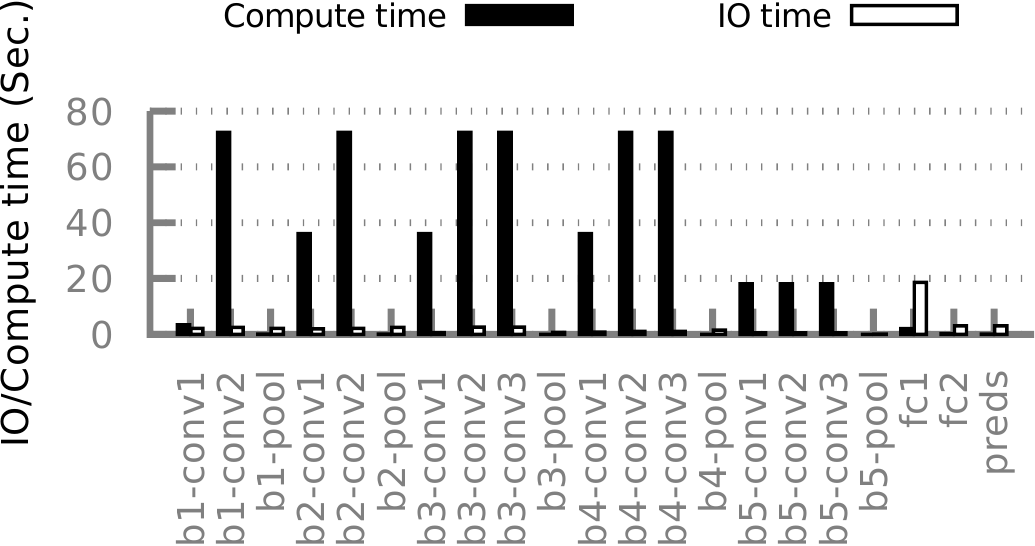}
        \caption{VGG16 (input shape: 224)}
        \label{fig:vgg-128k}
    \end{subfigure}
    ~ 
    \begin{subfigure}[b]{0.33\textwidth}
        \includegraphics[width=\textwidth]{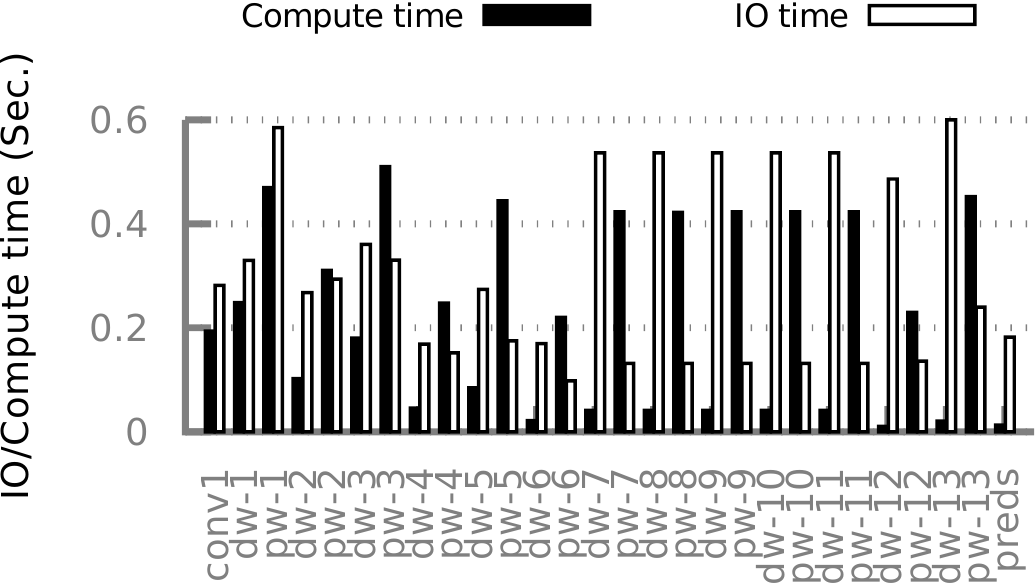}
        \caption{MobileNet (input shape: 224, alpha: 1)}
        \label{fig:mobilenet-224-1-128k}
    \end{subfigure}

    \caption{
    Per-layer compute and IO delays in NNs.
    (1) Observation:  NNs have a mix of IO-bound and compute-bound layers. (2) Insight: IO time can be hidden by compute time with parallel execution. 
    (3) Configuration: MCU is ARM Cortex-M7 @ 216 MHz, tile/buffer size is 128 KB, Transcend SD card size is 32 GB.}
    \label{fig:io-compute}
\end{figure*}

\begin{figure}[t!]
\centering
    \includegraphics[width=0.48\textwidth{}]{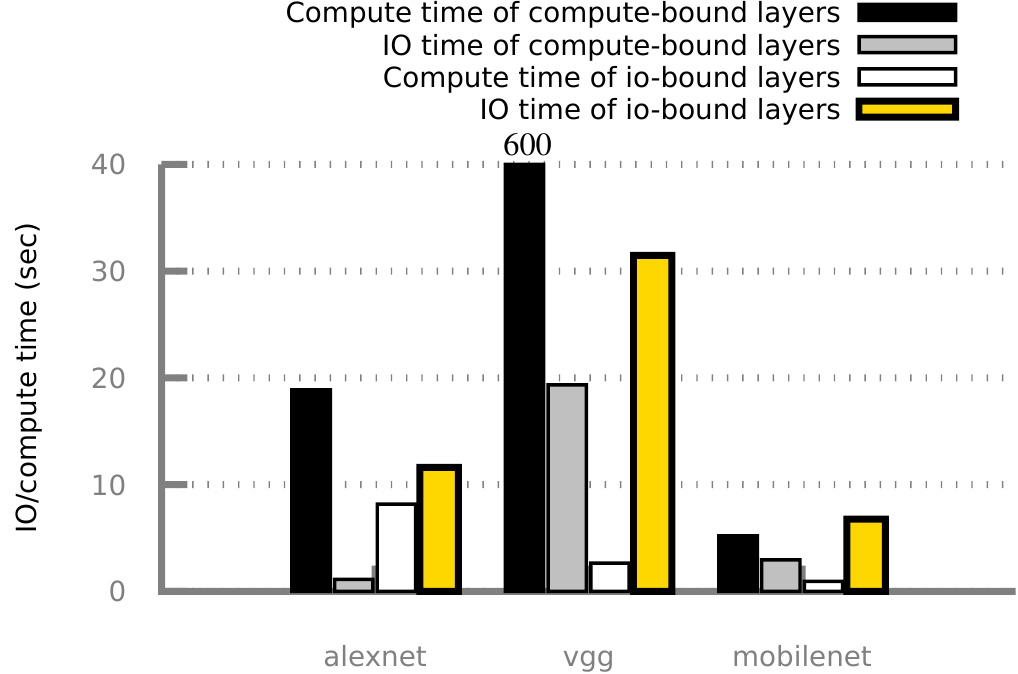}
    \caption{IO/compute delays in out-of-core NN execution. The total execution delay is dominated by compute in the compute-bound layers and IO in the IO-bound layers.
    }
    \label{fig:iob-cpb}
\end{figure}

To study the swapping overhead, 
we focus on a layer's swapping delay \textit{relative} to its computation delay on typical MCUs.  
The rationale is that as MCU can perform swapping and computation in parallel, the longer of the two delays will be the layer's bottleneck. 

\paragraph{Study setup}
Since the working set of a layer may not fit into SRAM, we split a layer's input, weight parameter, and output into small tiles (e.g., 128KB).
For compute time, we measure the time to calculate every output tile, then calculate the layer's compute time by adding every output tile's compute time.
For IO time, we measure the time to read input tiles, weight tiles (once), and output tiles, and then calculate the layer's IO time by adding them together.
Figure~\ref{fig:io-compute} shows the IO time and compute time of each layer in three typical CNNs,
where the buffer size for tiles is 128 KB.

\paragraph{Classifying NN layers}
In general, \textit{arithmetic intensity}, as commonly used in HPC ~\cite{roofline}, characterizes a workload's compute/IO ratio. It is defined as $W/Q$, where $Q$ is the amount of data to move in the memory hierarchy and $W$ is the amount of arithmetic operations on the data.
By factoring in an MCU's CPU speed ($S_{cpu}$) and IO bandwidth ($S_{IO}$), we define $N= (W/S_{CPU}) / (Q/S_{IO})$ as the \textit{normalized} arithmetic intensity on MCU. 
Of a given layer, 
$N > 1$ means swapping incurs less delay than computation, 
i.e, a compute-bound layer; 
$N < 1$ means swapping incurs longer delay, i.e. an IO-bound layer.

On modern MCUs with simple CPU cores, $S_{CPU}$ is primiarly determined by the CPU clockrate; 
it ranges from 64 MOPS to 480 MOPS~\cite{flops,cortexm}.
$S_{IO}$ is jointly determined by the MCU's DMA bandwidth and the SD card bandwidth, ranging from 10 MB/s to 40 MB/s as reported in literatures~\cite{microsdbw}. 
With these values, common NN layers fall into three distinct categories per their normalized arithemetic intensity ($N$). 

\noindent
\textit{(1) A majority of compute-bound layers} ($N >>1$). \hspace{2mm}
Notable examples are Conv layers known for their high complexity.
In the example of VGG16 (Table~\ref{tab:nn-layers-arithmetic-intensity}), 
$N$ for the Conv layers far exceeds 1 even with a high CPU clockrate and slow IO. 
They often dominate an NN's execution time (51\% -- 90\%), as exemplified by the three NNs in Figure~\ref{fig:io-compute}. 
On these layers, the computation delay overshadows the IO delay. 

\noindent
\textit{(2) Some IO-bound layers} ($N < 1$). \hspace{2mm}
Examples include fully connected (FC) and depth-wise convolutional layers (DW).
These layers perform light computation over large volumes of feature maps and weight parameters. 
Of all layers in an NN, they are often minorities (e.g. 2 out of 21 in VGG16). 
With out-of-core execution, the IO delay exceeds the computation delay by up to 10$\times$ (e.g. fc1 in Table~\ref{tab:nn-layers-arithmetic-intensity} and Figure~\ref{fig:vgg-128k}). 

\noindent
\textit{(3) Other layers with insignificant overheads}, e.g., Relu and Maxpooling.
These layers have low complexity and contribute a tiny fraction of data to move and to compute (0.3\%-0.9\%) for an NN. 
As such, their swapping overhead is insignificant. 

\paragraph{Common pattern of NN layers}
Based on the NN layer classification, there are two common patterns in typical CNNs:

(1) CNNs have a mix of compute-bound and IO-bound layers, and the number of compute-bound layers is usually larger than other layers.
Table~\ref{tab:cnn-study} shows the number of compute-bound and IO-bound layers in typical CNNs.
For instance, MobileNets~\cite{mobilenets}, Alexnet~\cite{alexnet}, VGG16~\cite{vgg}, ResNet18~\cite{resnet}, and GoogLeNet~\cite{googlenet} have 14/13, 5/3, 13/2, 16/2, and 21/6 of compute-bound/IO-bound layers respectively.

(2) The overall CNN execution time is dominated by the compute time of compute-bound layers and the IO time of IO-bound layers.
Figure~\ref{fig:iob-cpb} shows the IO time and compute of IO-bound/compute-bound layers.
For instance, compute time of compute-bound layers dominate the overall time in Alexnet and VGG. 
For Mobilenet, the IO-time of IO-bound layers dominates the overall time, because Mobilenet is using specially point-wise and depth-wise convolutions~\cite{mobilenets}, which have lower compute complexity than general convolutional layers.

\textbf{\textit{Insights}}: \textit{Towards lowering the swapping overhead, we exploit the aforementioned pattern of NN layers. By executing compute-bound layers and IO-bound layers in parallel, we hide the IO delays behind the compute delays.
}

\subsection{The system model}
\label{sec:model}

\paragraph{MCU hardware}
We assume the following hardware components:
(1) a CPU with clockrate from tens of MHz to a few hundred MHz, as exemplified by Arm Cortex M3 and M7; 
(2) on-chip SRAM: from tens of KBs to several MBs; 
(3) on-chip NOR flash: byte-addressable, read-only memory no more than a few MBs;
(4) cheap external storage, e.g. a micro SD card ranging from tens of GBs to a few hundred GBs; 
(5) a DMA engine, for moving data between SRAM and external storage without CPU involved;  
(6) optionally, on-chip accelerators for computing crypto and hash functions. 

Major vendors ship numerous MCU models meeting the above conditions. 
Examples include the STM32 MCU family from STMicroelectronics~\cite{stmcu} and the LCP series from NXP Semiconductors~\cite{nxpmcu}. 
They are priced at \$1-\$20 per unit. 


\paragraph{NN workloads \& metrics}
We motivate our study by considering periodic NN inference on video/audio data as a sequence of \textit{frames} captured by MCUs at run time. 
To characterize inference speed, we consider both the inference delay of each frame and throughput as the number of frames processed per second. 
MCU applications may be sensitive to either metric or both. 
For instances, keyword spotting is sensitive to inference delays~\cite{zhang2017hello} 
and car counting benefits from high throughput~\cite{xu2020approximate}.

\begin{figure}[t!]
\centering
    \includegraphics[width=0.45\textwidth{}]{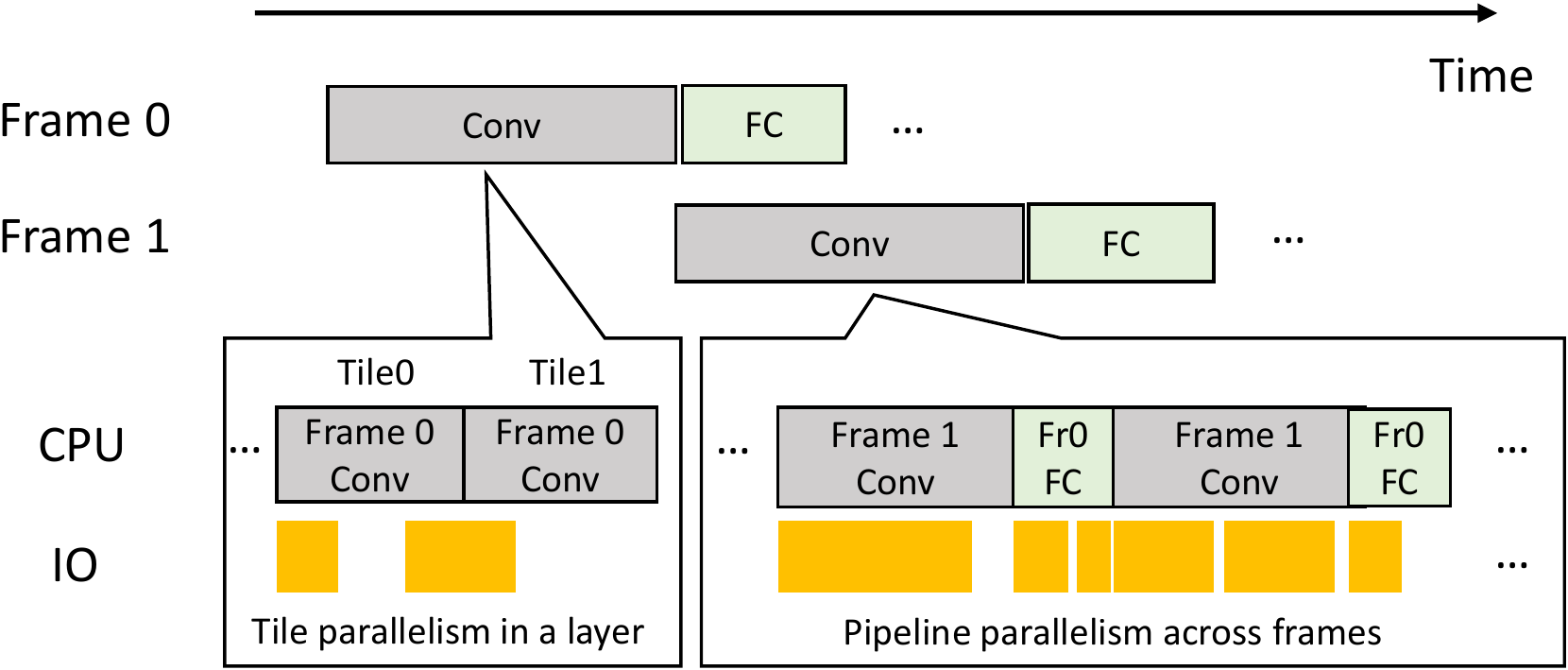} 
    \caption{An example of out-of-core NN execution, showing Conv (compute-bound) and FC (IO-bound) layers.}
    \label{fig:swap-model}
\end{figure}

\paragraph{Out-of-core NN executions}
We consider the following swapping strategy.
An NN's parameters are pre-stored on the external flash. 
Given an input frame, the MCU executes the NN's layers in sequence.
It processes a layer in tiles, in case the layer's memory footprint exceeds MCU's main memory: 
to do so, the MCU loads to the main memory a tile of parameters and a tile of input feature maps, computes a tile of output feature maps in memory, and writes back the output to the external flash. 
Altogether, the input and output tiles shall simultaneously fit in the main memory. 

As shown in Figure~\ref{fig:swap-model}, 
MCU extracts CPU/IO parallelism for hiding IO delays.
(1) \textit{Tile parallelism within an NN layer}: 
while computing an output tile \textit{Tile0}, MCU can pre-load from flash the input tiles for computing the next output tile \textit{Tile1}; 
while writing back the completed \textit{Tile0} back to flash, MCU can compute \textit{Tile1} simultaneously. 
(2) \textit{Layer parallelism}: 
in a similar fashion, MCU can execute an earlier layer's computation with a latter layer's IO simultaneously. 
(3) \textit{Pipeline parallelism across data frames}: 
MCU can execute compute-bound and IO-bound layers for different frames in parallel, as these layers exercise complementary resources, namely CPU and IO bandwidth. 
As shown in Figure~\ref{fig:swap-model}, MCU swaps frame 0's FC layer while computes on frame 1's Conv layer.

\section{\sys{}: Automatically scheduling IO/Compute tasks in parallel}

In order to reduce IO overhead in swapping, we present \sys{}, a scheduler design that automatically schedules IO tasks and compute tasks across tiles, layers and frames in parallel based on NN characteristics, meanwhile respects memory constraint and data dependency.

\subsection{Challenges}
As shown in Figure~\ref{fig:swap-model}, MCU ideally could extract CPU/IO parallelism for hiding IO delays.
However, such ideal parallel scheduling sequence is difficult to find because it must meet the following requirements at the same time:
(1) the scheduler must automatically identify what tiles should be executed in parallel according to their dependencies and relative IO/compute time;
(2) the working set of tiles being executed in parallel must be smaller than SRAM at every single moment;
(3) the parallel sequence should keep both MCU core and IO bandwidth fully utilized to avoid either of them from idling.

Furthermore, divers NN layers with different parameters and diverse SRAM sizes on MCUs create a huge space of choices for deciding parallel sequence for IO/compute tasks, which makes parallel scheduling even more difficult.


\subsection{\sys{} design}
To address the above challenges, we present the design of \sys{}, describing how to decide tile size, manage memory buffers, and schedule IO/compute tasks in parallel, meanwhile respect memory constraint, data dependency, and task priority.
\begin{figure*}
\centering
    \includegraphics[width=0.95\textwidth{}]{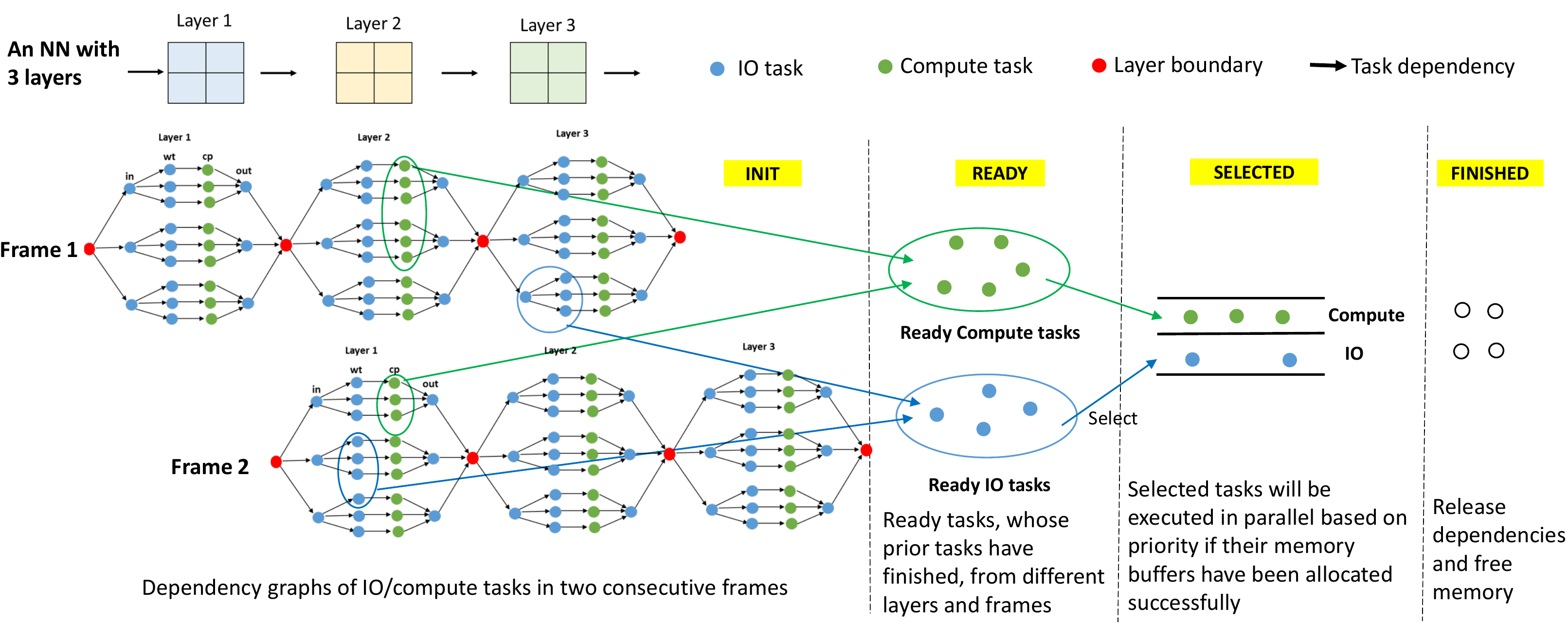} 
    \caption{Overview of \sys{}: scheduling IO/compute tasks across tiles, layers, and frames in parallel according to dependencies, priorities, and memory constraints.}
    \label{fig:scheduler}
\end{figure*}

\paragraph{Tiling NN layers and managing memory buffers}
A key question in swapping is to decide tile sizes for NN layers based on SRAM size.
\sys{} splits SRAM size into fixed number of buffers, and then calculates tile sizes based on layer parameters and buffer size.
Specifically, the input tile size depends on the output tile size, so they will be decided together and the larger one of them must be smaller than buffer size. Weight tile size doesn't depend on input or output, so it is calculated just according to weight size and buffer size.

As show in Algorithm~\ref{algo:scheduler}, \sys{} equally splits SRAM into buffers with fixed size, and creates three separate memory buffer pools for input feature maps, weight parameters, and output feature maps, who have 1/4, 1/2, and 1/4 of total memory buffers.
The reason why \sys{} creates separate memory pools, instead of one pool, is that single memory pool for input/weight/output tiles leads to deadlock in parallel execution.
For example,  all memory buffers may be allocated to input and weight tiles, so execution cannot continue because of no memory buffers for output tiles.
The rational to choose 1/4, 1/2, and 1/4 is based on the minimal parallel working set of computing one output tile, which includes one input tile, at least two weight tiles, and one output tile.


\paragraph{NN task and graph} As shown in Figure~\ref{fig:scheduler}, \sys{} defines two types of tasks: IO task and compute task.
An IO task reads/writes tiles from/to SD, and a compute task computes an output tile based on corresponding input/weight tiles.

\sys{} defines an NN as a computation graph \textit{G = (V, E)}, 
where \textit{V} is the node set of IO and compute tasks, and \textit{E} is the edge set representing dependencies.
For instance, a compute task depends on IO tasks that read input/weight tiles, and a write IO task depends on a compute task that finishes computing output tile. 
Every task has a set of properties, e.g., \textit{in-degree} counter indicating the number predecessors of current task, memory buffer and tile sizes, execution time, and execution priority.

Two things that are worth noting in NN graph: (1) we enforce dependencies between an input tile and multiple weight tiles to ensuring reading input tile first, 
so that reading other weight tiles can happen in parallel with computing an output tile.
(2) each output tile depends on all weight tiles, so weight tiles may be read multiple times (once for each output tile) during execution. 
 
As show in Algorithm~\ref{algo:scheduler}, \textit{BuildGraph()} takes NN architecture and SRAM size as parameters. 
For each layer, \sys{}: 
(1) calculates tile sizes for input/weight/output based on memory buffer size;
(2) creates read IO tasks for input and weight tiles, compute tasks for computing output tiles, and write IO tasks for output tiles;
(3) inserts IO/compute tasks to execution graph based on dependencies;
(4) sets task properties, including execution time, memory buffer size, \textit{inDegree} counter, and priority.

\paragraph{Task state}
As shown in Figure~\ref{fig:scheduler}, \sys{} defines the following states for every IO/compute task to manage their lifecycle:
\begin{itemize}
	\item \textbf{INIT} A task is set to INIT state when building the execution graph based on NN architecture, layer parameters, SRAM size, buffer size, and dependency.
	 
	\item \textbf{READY} A task becomes READY when all of its predecessors have finished, at which point the \textit{in-degree counter} of the task drops to zero.
	
	  
	\item \textbf{SELECTED} 
	A task switches to SELECTECD from READY when its memory buffers has been successfully allocated, e.g., an input/weight buffer for a read IO task or an output buffer for a compute task.
	
	\item \textbf{FINISHED} When a IO/compute task is finished, it switches to FINISHED state, at which point \sys{} decreases the \textit{in-degree counter} by one for the task's all successors to release the dependency and free memory buffers accordingly.
	
\end{itemize}

\paragraph{Task priority}
When there are multiple READY tasks from multiple layers and frames, the tasks from earlier frames/layers should have higher priority to be executed to guarantee per frame delay.
\sys{} assigns priority to tasks based on their frame number and layer number when creating these tasks, and schedules them at runtime according to the priority.

\paragraph{Scheduling NN tasks}
Given tiling strategies of NN layers, \sys{} finds the optimal parallel sequences for IO and compute tasks based on their dependencies, available memory buffers, and priority.
One goal of scheduling is to keep both MCU and IO busy to avoid either of them from idling, to achieve low latency and high throughput.

As show in Algorithm~\ref{algo:scheduler}, \sys{} maintains two tasks queues, \textit{ReadyIO} and \textit{ReadyCP}, for READY IO tasks and READY compute tasks respectively.
READY tasks in these two queues are sorted based on their priority, and the one with the highest priority will be scheduled each time.

\textit{ScheduleIOTask()} keeps looking for IO tasks in ReadyIO queue in priority order. 
For write IO tasks that do not require memory allocation, \sys{} issues write DMA operation, and then frees memory buffers and releases the dependencies for the task's successors.
For read IO task, \sys{} first tries to allocate memory buffer for it. If the allocation succeeds, then issue read DMA operation and release the dependencies for the task's successors.

\textit{ScheduleComputeTask()} keeps looking for compute tasks in ReadyCP queue in priority order. It first tries to allocate memory buffers to store computing output. If the allocation succeeds, \sys{} executes the compute task, release the dependencies for its successors, and free memory buffers of input/weight tiles.

With two separate threads running \textit{ScheduleIOTask()} and \textit{ScheduleComputeTask()}, 
\sys{} can schedule any ready IO and compute tasks in parallel across tiles, NN layers, and data frames, meanwhile respects memory constraint, data dependency, and task priority. 
Therefor, the IO overhead in swapping can be reduced.



\begin{algorithm}
\small
\SetKwInOut{Input}{Input}
\SetKwInOut{Output}{Output}
\Input{NN architecture and SRAM size}


$Layers[L]$ = parameters of L layers in an NN\\

$ReadyIO$ = a set of READY IO tasks sorted by priority \\
$ReadyCP$ = a set of READY Compute tasks sorted by priority\\


\SetKwFunction{FBuildGraph}{BuildGraph}
    \SetKwProg{Fn}{Function}{:}{}
    \Fn{\FBuildGraph{$Layers$, $SRAMSize$}}{
	    $G$ = empty graph \\
	    \For{$Layer \in Layers$} {
			Calculate tile sizes for input, weight, and output; \\
		    \For{all input tiles} {
				$Insert$ an IO task for reading input tile to G; \\
				\For{all weight tiles} {
					$Insert$ an IO task for reading weight tile to G; \\
					$Insert$ a Compute task to G; \\
				}
				\textit{Insert} an IO task for writing output tile to G;
		    }
		    
	    }
	    Set the root IO Task to READY and insert into $ReadyIO$; \\
        \textbf{return} $G;$ 
}

	

\SetKwFunction{FScheduleIOTask}{ScheduleIOTask} 
	\SetKwProg{Fn}{Function}{:}{}
	\Fn{\FScheduleIOTask{$G$}}{
	\While{$readyIO$ is NOT empty()} {
		\For (\tcp*[h]{in priority order}){iotask in $ReadyIO$} {
			\If{iotask is WRITE} {
				Execute the write iotask; \\
				ReleaseSuccessors($G$, $iotask$); \\
				freeMemoy(iotask.buffptr); \\
			}
			\Else(\tcp*[h]{iotask is READ}){
				buffptr = AllocateMemory(); \\
				\If{buffptr != NULL} {
					Execute the READ IO task;\\
					ReleaseSuccessors($G$, $iotask$); \\
				}
			}
		}
	}
}
\texttt{\\} 

\SetKwFunction{FScheduleComputeTask}{ScheduleComputeTask} 
	\SetKwProg{Fn}{Function}{:}{}
	\Fn{\FScheduleComputeTask{$G$}}{
	\While{$readyCP$ is NOT empty()} {
		\For (\tcp*[h]{in priority order}){cptask in $ReadyCP$} {
			buffptr = AllocateMemory(); // for output tile\\
			\If{buffptr != NULL} {
				Execute the Compute task;\\
				ReleaseSuccessors($G$, $cptask$); \\
				freeMemory(); // for input and weight tiles 
			}
		}
	}
}
\texttt{\\} 

\SetKwFunction{FReleaseSuccessors}{ReleaseSuccessors} 
	\SetKwProg{Fn}{Function}{:}{}
	\Fn{\FReleaseSuccessors{$G$, $task$}}{
		\For{$suctask$ in $task$'s successors} {
			\If{$suctask$.inDegree - - == 0} {
				insert $suctask$ to ReadyIO or ReadyCP;
			}
		}
	}
\texttt{\\} 

$freeListIn, freeListWt, freeListOut$ = lists of free memory buffers for input, weight, and output tiles\\
\SetKwFunction{FAllocateMemory}{AllocateMemory} 
	\SetKwProg{Fn}{Function}{:}{}
	\Fn{\FAllocateMemory{$freeList$}}{
		\If{$freeList$ is empty} {
			return NULL;
		}
		buffptr = select one buffer from $freeList$; \\
		return buffptr;
	}

\SetKwFunction{FFreeMemory}{FreeMemory} 
	\SetKwProg{Fn}{Function}{:}{}
	\Fn{\FFreeMemory{$buffptr$, $freeList$}}{
		insert $buffptr$ to $freeList$;
	}

\caption{Scheduling IO/compute tasks in parallel}
\label{algo:scheduler}
\end{algorithm}

\section{Implementation \& Methodology}

\paragraph{Implementation}
We implement swapping kernels for typical NN layers to compute tiles on MCU atop CMSIS-NN library~\cite{cmsisnn}, and currently supported layers include Convolution, ReLu, Pooling, Fully Connected, Depth-wise convolution, and Point-wise convolution.
We implement the scheduler in C++, which can run on desktop to find the best parallel scheduling sequence without deploying on MCUs.

\paragraph{Studied NNs}
We study three representative NNs, whose memory footprints range from sveral-MB to hundred-MB (with quantization).
As shown in Table~\ref{tab:cnn-study}:
MobileNet has large feature maps but small weight parameters,
AlexNet has small feature maps but large weight parameters,
and VGG16 has 1000$\times$ larger memory footprint than MCUs' SRAM size.

\paragraph{Input data}
We use synthetic images as the input. Note that the input contents do not affect NN execution
time/efficiency, hence our measurement results.

\paragraph{Methodology}
In order to understand how swapping affects the latency, throughput, SD durability, energy consumption, and security, we do the following steps for all three NNs:
(1) Given SRAM size and buffer size, calculate the tile sizes for all layers of an NN;
(2) Based on tile sizes of layers, we run the swapping kernels as microbenchmarks on target MCU hardware (TM32F746NG-Discovery board: ARM Cortex-M7 at 216 MHz, 340 KB SRAM, 32 GB SD card), and then measure the IO/compute time for tiles;
(3) The scheduler takes NN architecture/parameters, SRAM size, buffer/tile sizes, IO/compute time of tiles as parameters, and then automatically finds out the optimal parallel scheduling sequences for IO and compute tasks across layers and frames. 

For latency, we measure the time to process one NN frame. 
For throughput, we measure the time to process 10 consecutive NN frames in parallel and then calculate the throughput. 
For energy, we measure the \textit{worst-case} energy consumption by keep running IO and compute tasks simultaneously.

\section{Findings}

\begin{figure*}
    \centering
    \begin{subfigure}[b]{0.24\textwidth}
        \includegraphics[width=\textwidth]{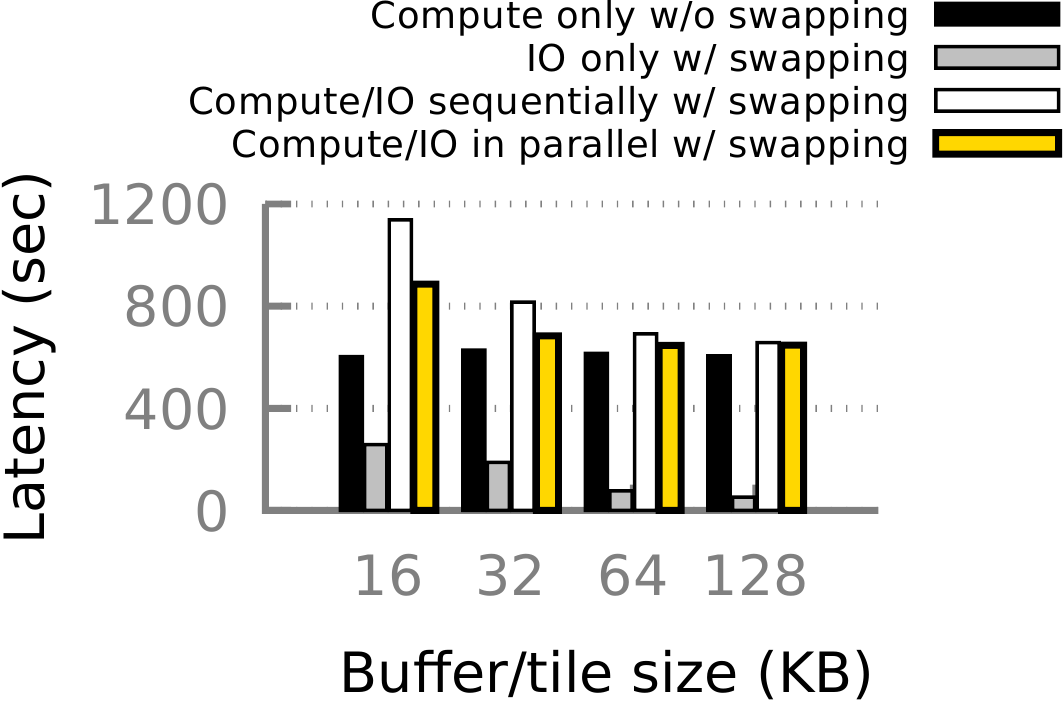}
        \caption{VGG, SRAM 512KB}
        \label{fig:vgg-512k}
    \end{subfigure}
    ~ 
    \begin{subfigure}[b]{0.24\textwidth}
        \includegraphics[width=\textwidth]{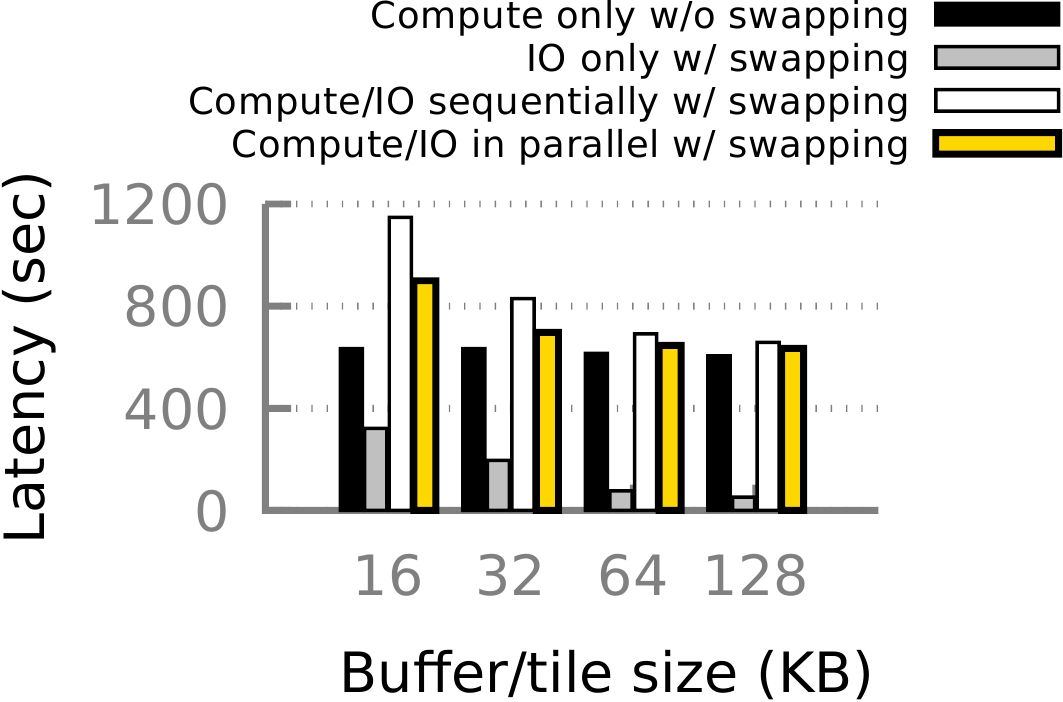}
        \caption{VGG, SRAM 1MB}
        \label{fig:vgg-1m}
    \end{subfigure}
    ~ 
    \begin{subfigure}[b]{0.24\textwidth}
        \includegraphics[width=\textwidth]{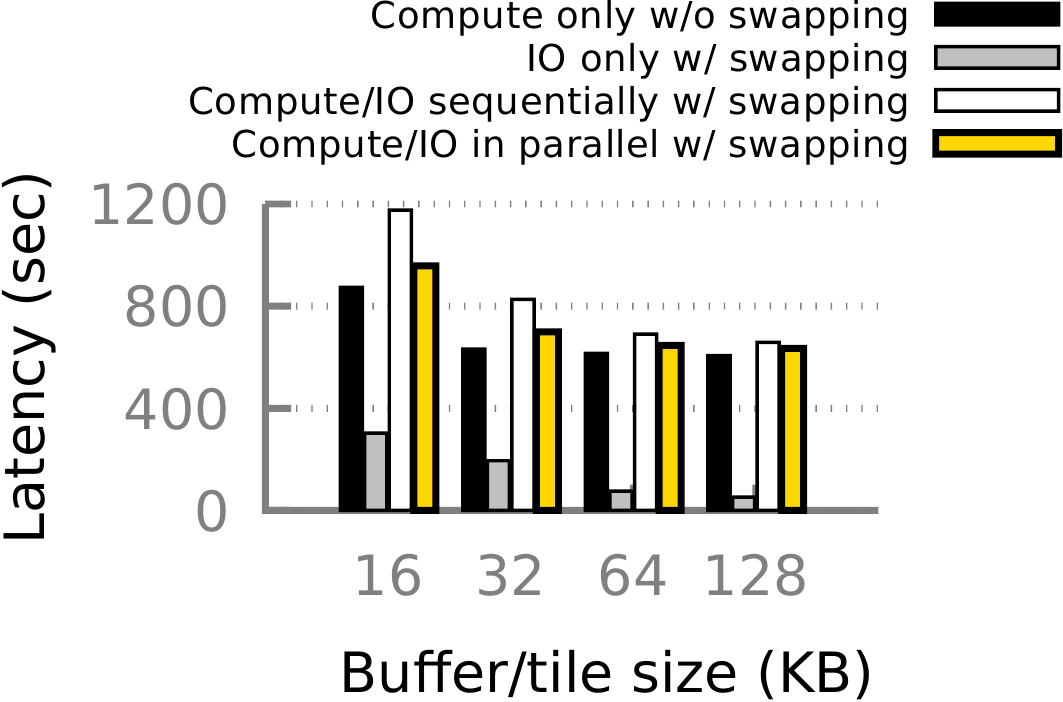}
        \caption{VGG, SRAM 4MB}
        \label{fig:vgg-4m}
    \end{subfigure}
    ~ 
    \begin{subfigure}[b]{0.24\textwidth}
        \includegraphics[width=\textwidth]{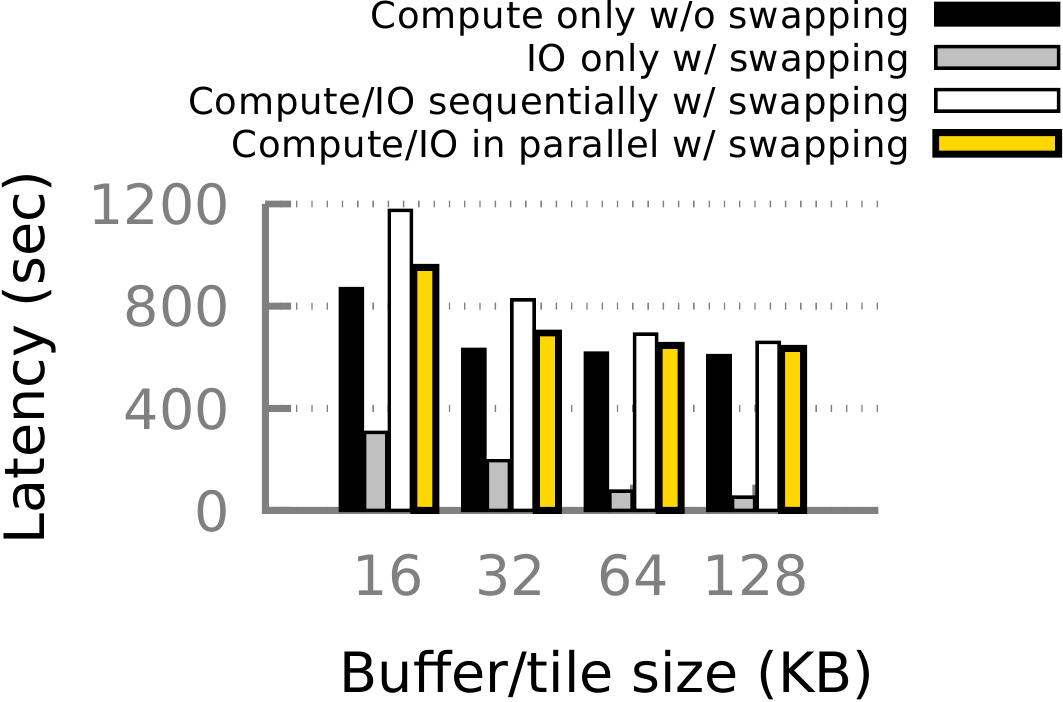}
        \caption{VGG, SRAM 8MB}
        \label{fig:vgg-8m}
    \end{subfigure}
    

    \begin{subfigure}[b]{0.24\textwidth}
        \includegraphics[width=\textwidth]{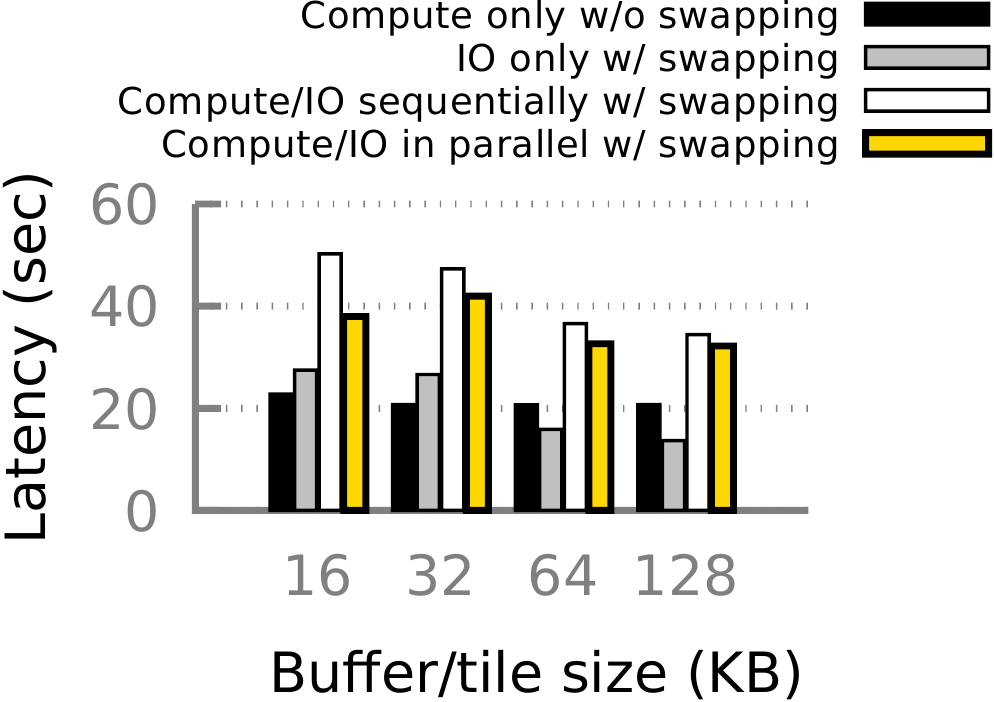}
        \caption{AlexNet, SRAM 512KB}
        \label{fig:alexnet-512k}
    \end{subfigure}
    ~ 
    \begin{subfigure}[b]{0.24\textwidth}
        \includegraphics[width=\textwidth]{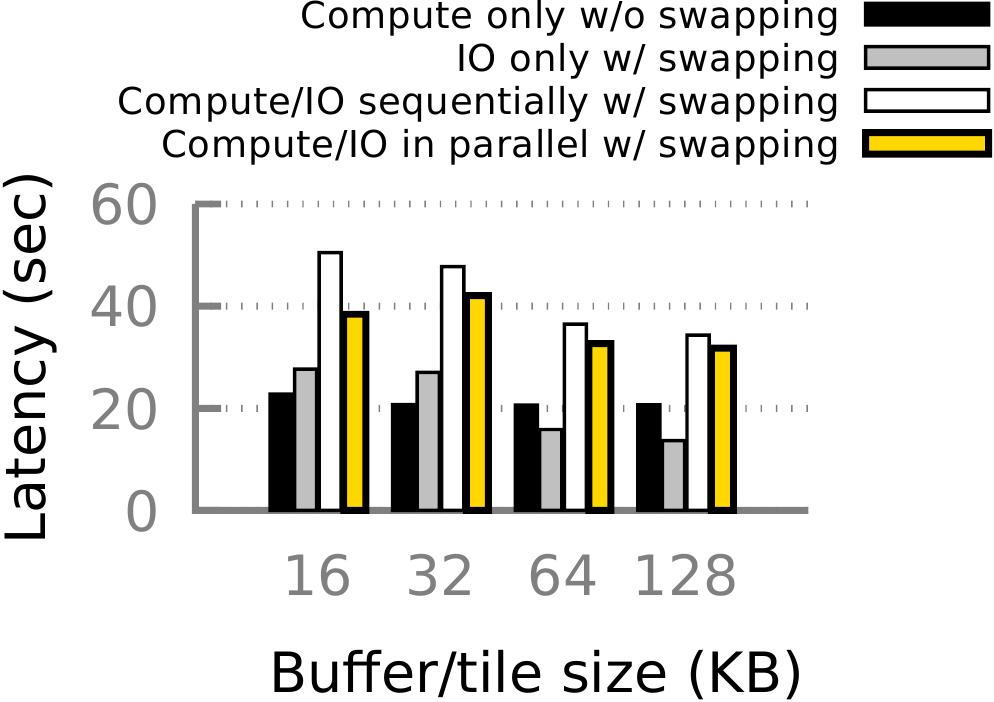}
        \caption{AlexNet, SRAM 1MB}
        \label{fig:alexnet-1m}
    \end{subfigure}
    ~ 
    \begin{subfigure}[b]{0.24\textwidth}
        \includegraphics[width=\textwidth]{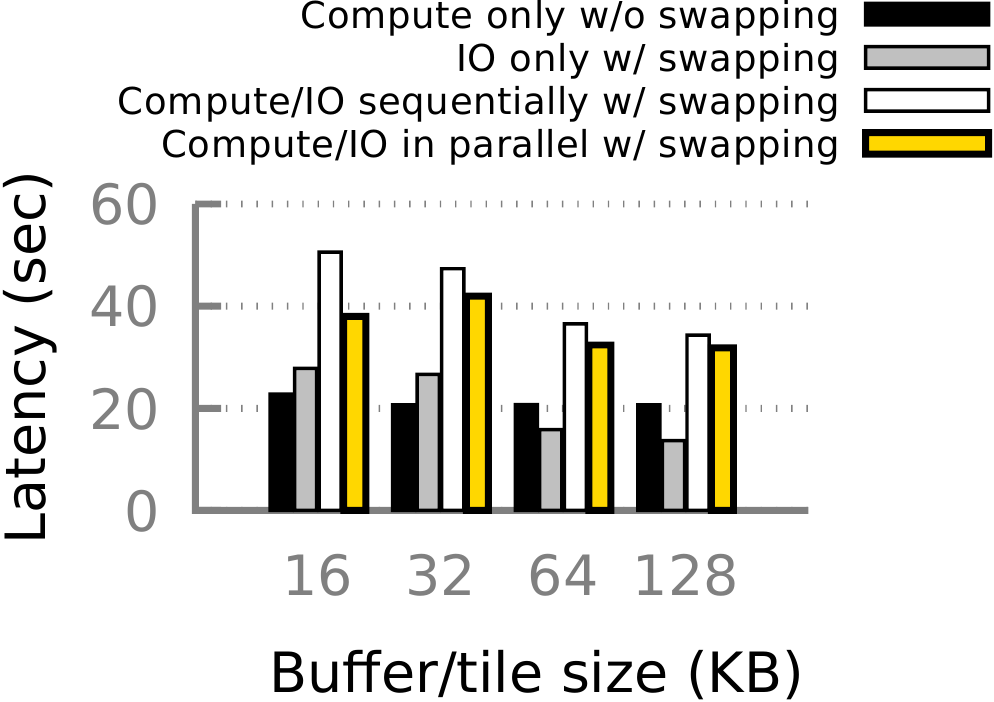}
        \caption{AlexNet, SRAM 4MB}
        \label{fig:alexnet-4m}
    \end{subfigure}
    ~ 
    \begin{subfigure}[b]{0.24\textwidth}
        \includegraphics[width=\textwidth]{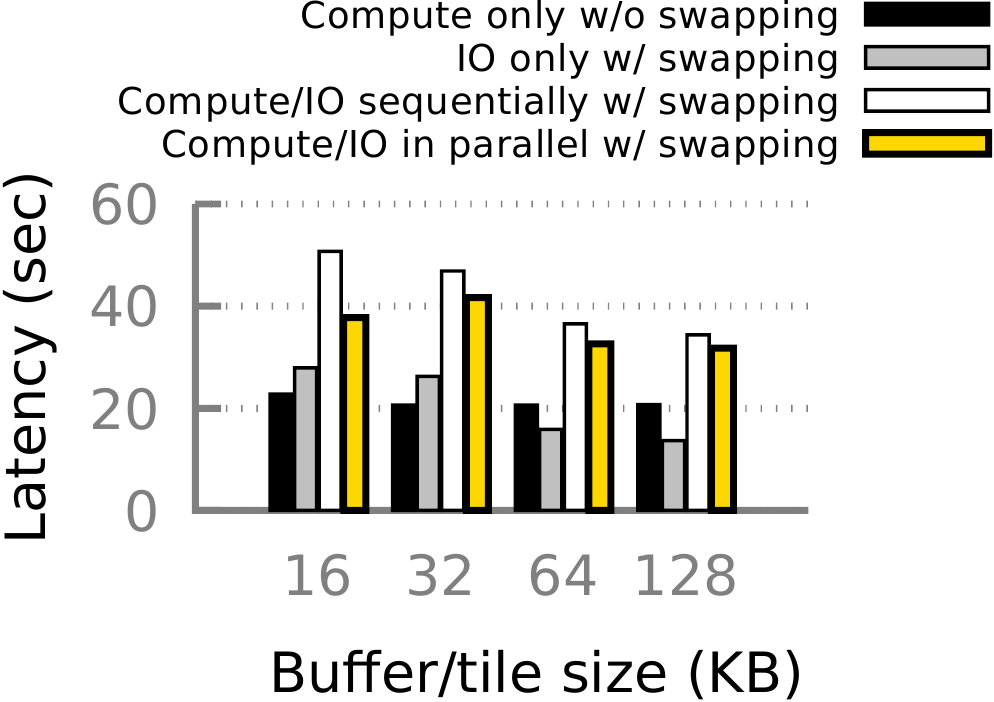}
        \caption{AlexNet, SRAM 8MB}
        \label{fig:alexnet-8m}
    \end{subfigure}
    
    
    \begin{subfigure}[b]{0.24\textwidth}
        \includegraphics[width=\textwidth]{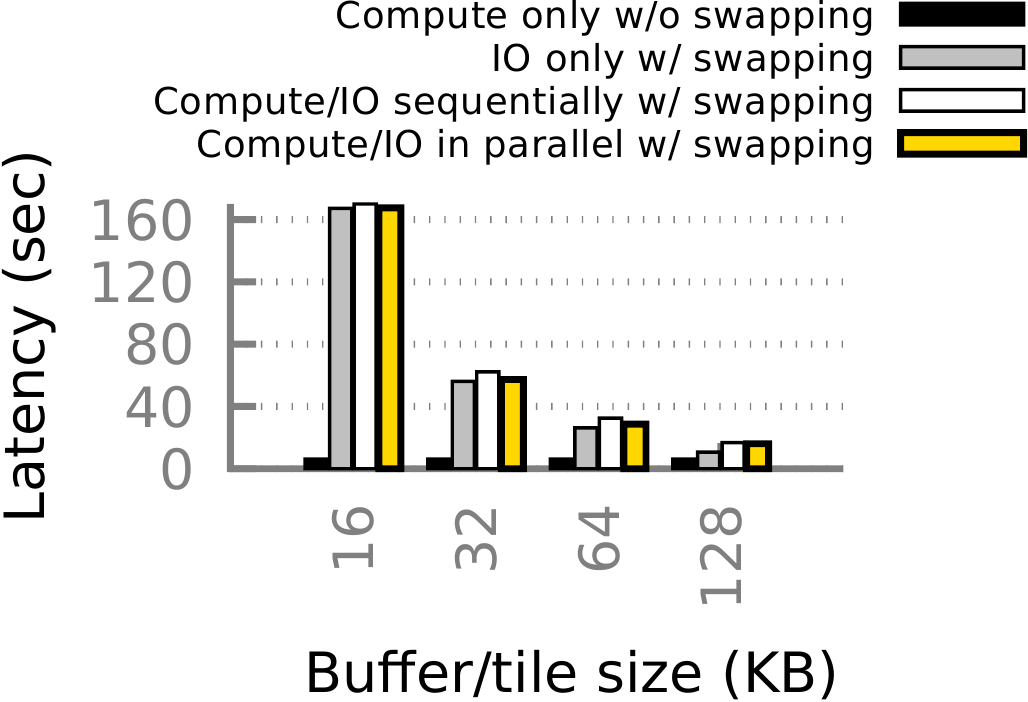}
        \caption{Mobilenet, SRAM 512KB}
        \label{fig:mobilenet-512k}
    \end{subfigure}
    ~ 
    \begin{subfigure}[b]{0.24\textwidth}
        \includegraphics[width=\textwidth]{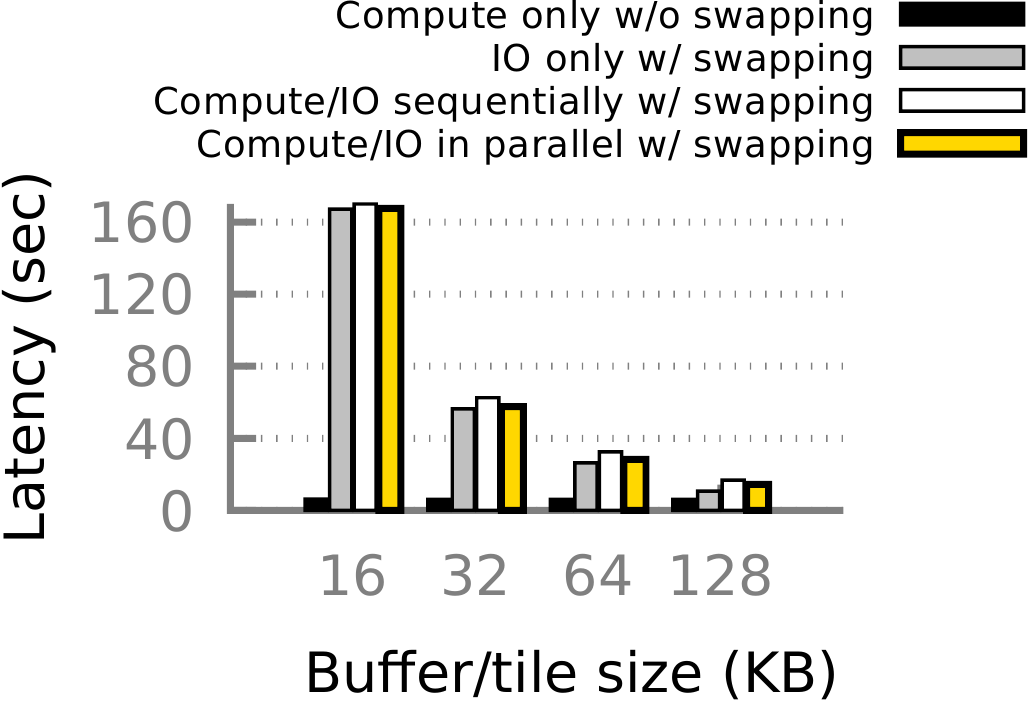}
        \caption{Mobilenet, SRAM 1MB}
        \label{fig:mobilenet-1m}
    \end{subfigure}
    ~ 
    \begin{subfigure}[b]{0.24\textwidth}
        \includegraphics[width=\textwidth]{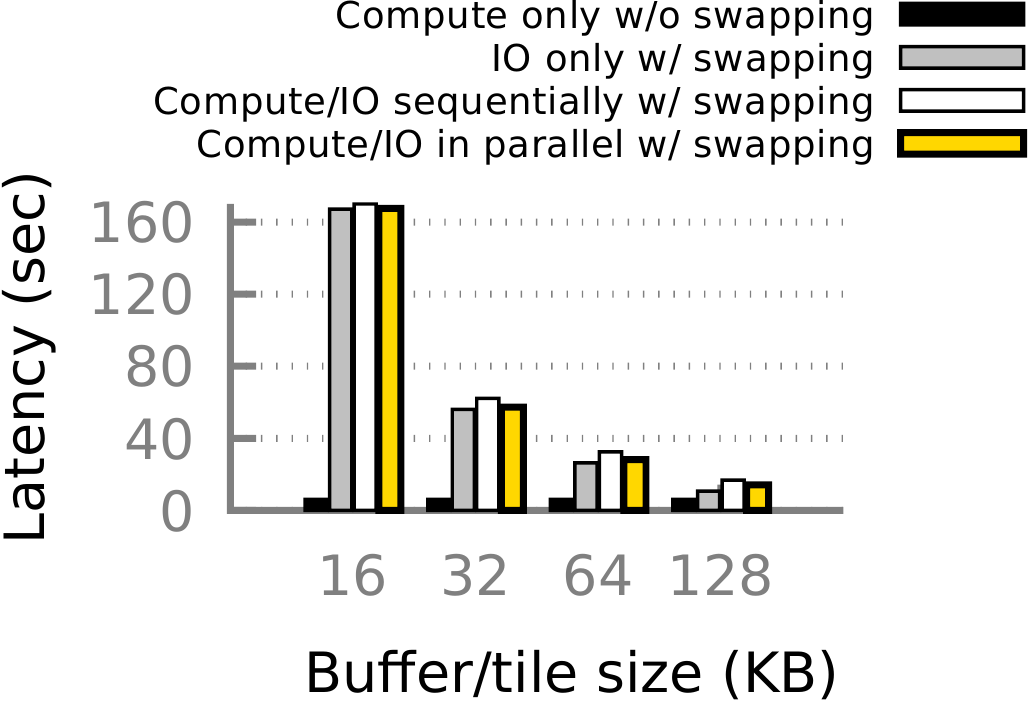}
        \caption{Mobilenet, SRAM 4MB}
        \label{fig:mobilenet-4m}
    \end{subfigure}
    ~ 
    \begin{subfigure}[b]{0.24\textwidth}
        \includegraphics[width=\textwidth]{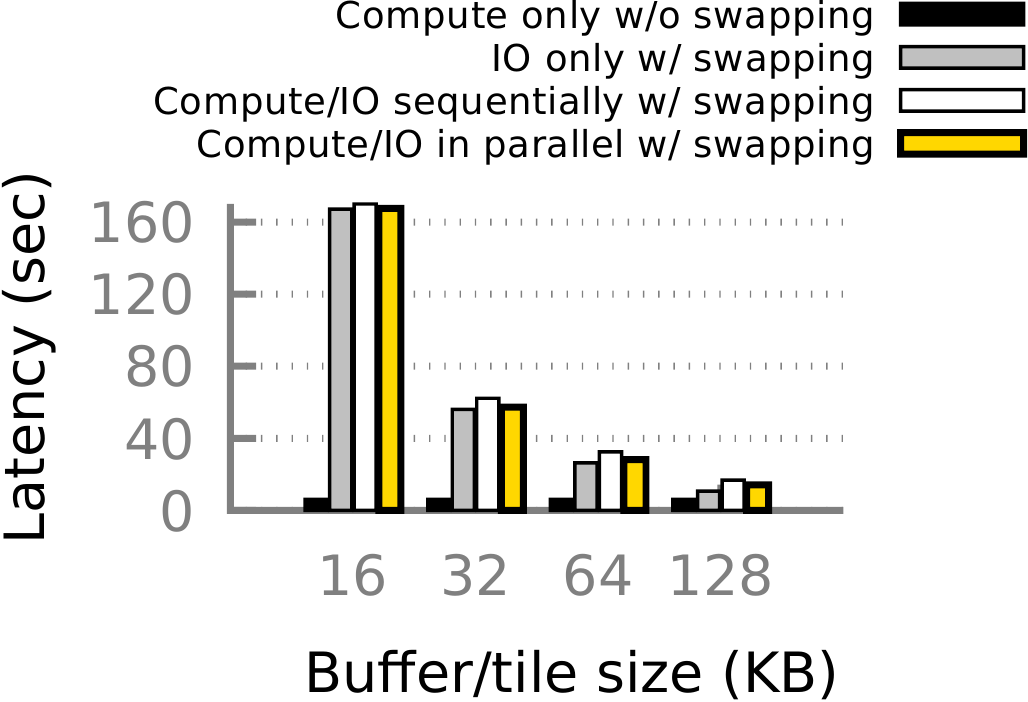}
        \caption{Mobilenet, SRAM 8MB}
        \label{fig:mobilenet-8m}
    \end{subfigure}

    \caption{Swapping latency of NNs with different SRAM sizes and buffer sizes. 
    Observation: 
    swapping incurs negligible or modest delay in latency.
    }
    \label{fig:swapping-latency}
\end{figure*}

\begin{figure*}
    \centering
    \begin{subfigure}[b]{0.3\textwidth}
	   \includegraphics[width=\textwidth]{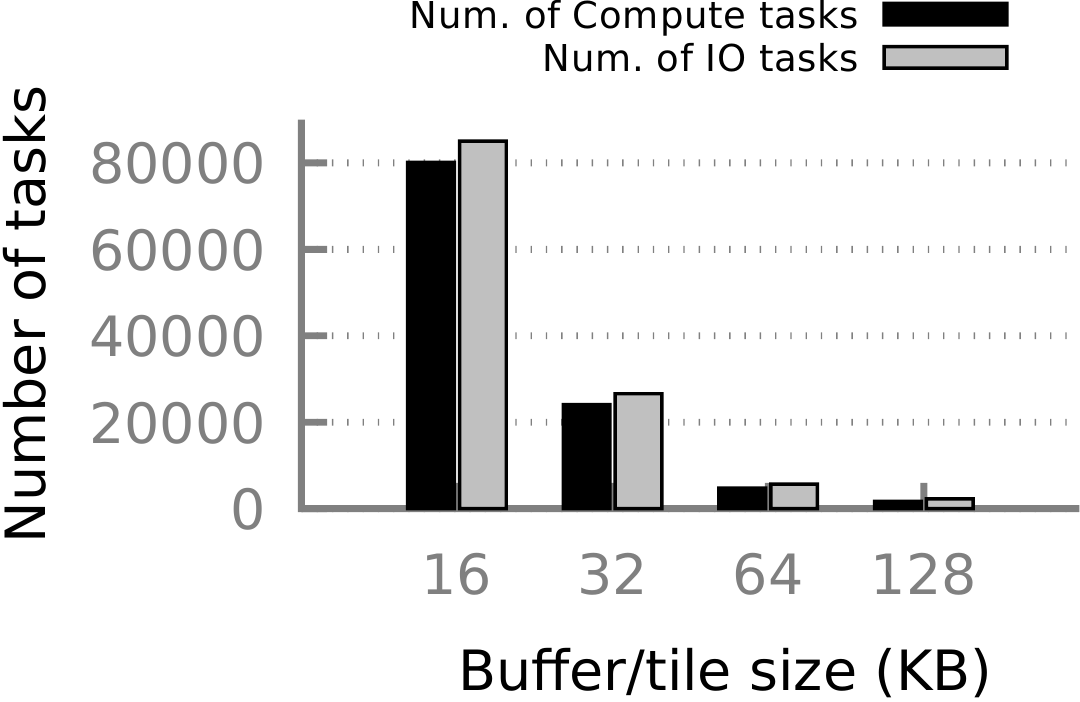}
       \caption{VGG}
       \label{fig:vgg-tasks}
    \end{subfigure}
    ~
    \begin{subfigure}[b]{0.3\textwidth}
        \includegraphics[width=\textwidth]{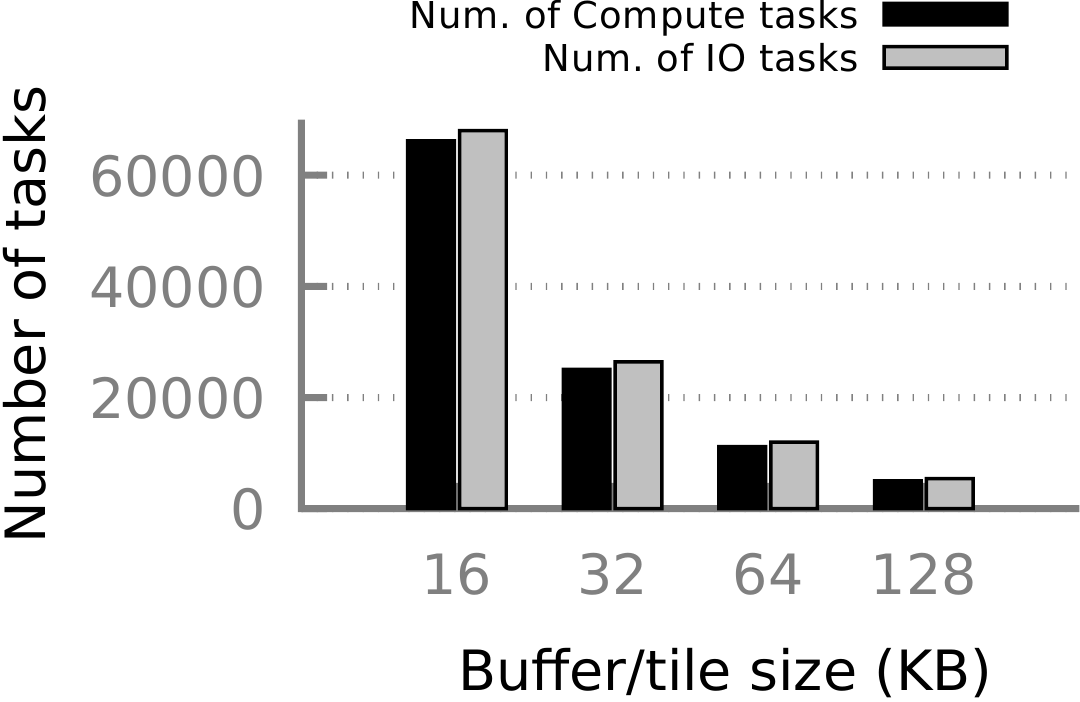}
        \caption{AlexNet}
        \label{fig:alexnet-tasks}
    \end{subfigure}
    ~ 
    \begin{subfigure}[b]{0.3\textwidth}
        \includegraphics[width=\textwidth]{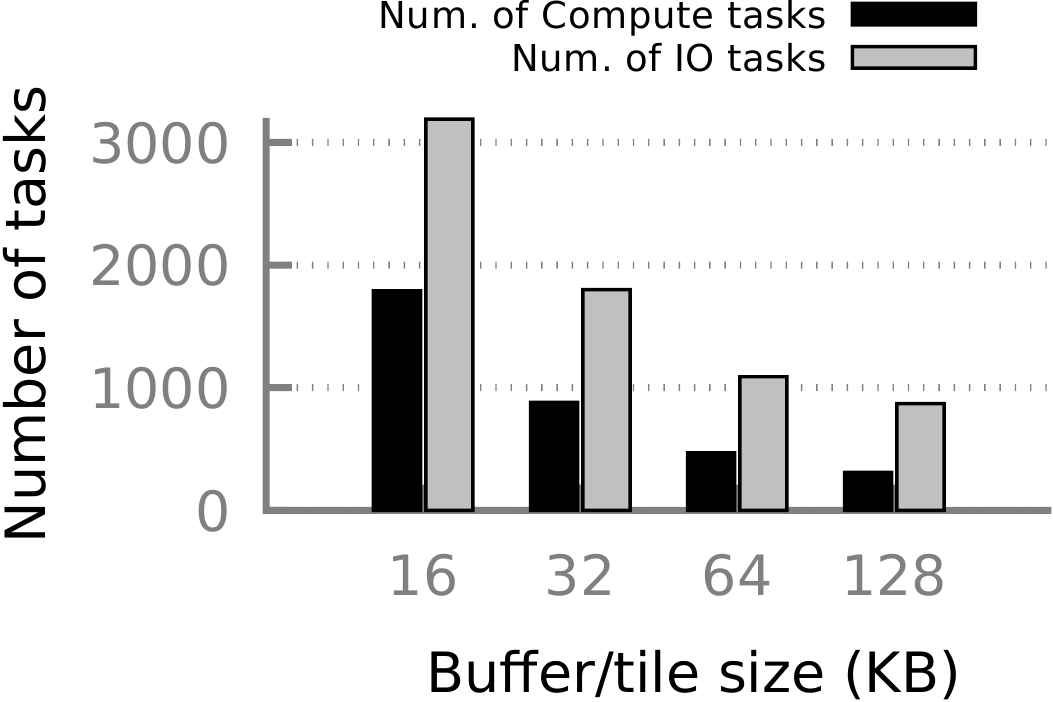}
        \caption{Mobilenet}
        \label{fig:mobilenet-tasks}
    \end{subfigure}

    \caption{Number of IO/compute tasks in NNs under different buffer/tile sizes. Observation: the number of IO/compute tasks drops significantly as the buffer/tile size increases.}
    \label{fig:swapping-tasks}
\end{figure*}

This section focus on the analysis and findings of out-of-core NN on MCUs by answering the following questions:
\begin{itemize}
	\item What are the parameters and tradeoffs that affect swapping performance?
	\item How does swapping affect per-frame latency?
	\item How does swapping affect throughput?
	\item Will swapping wear out SD soon?
	\item How much extra energy does swapping consume?
	\item Does swapping incur security issues?
\end{itemize}

\subsection{Software/hardware parameters and their tradeoffs}
There are multiple hardware/software parameters that affect the swapping performance, 
including SRAM size, buffer/tile size, the number of buffers, NN's memory footprint, and the ratio of compute-bound and IO-bound layers in NNs. We analyze each of them as following:
\begin{itemize}

	\item \textit{SRAM size:} Large SRAM leads to large memory buffers or more memory buffers, but also increases cost and energy consumption. 

	\item \textit{Buffer/tile size:} Tile is a small chunk of input/weight/output, and it decides the granularity of IO/compute task.
	Small tiles lead to fine-grained tasks and therefore better compute/IO parallelism, but they increase the total amount of IO traffic/time.
	Buffers are used to store tiles, and tile size is calculated based on buffer size. We treat them the same in discussion.

	\item \textit{The number of memory buffers:} The more, the better. More memory buffers allows more tiles co-existing in SRAM, so more tasks can be executed in parallel.
 
	\item \textit{NN's memory footprint:} It's decided by NN architecture. 
	NNs with larger memory footprint see higher IO overhead in swapping due to more IO traffic, and vice versa.

	\item \textit{The ratio of compute-bound and IO-bound layers in NNs}. It'd decided by NN architecture and affects the IO overhead in swapping. NNs with more compute-bound layers, the IO overhead is lower since IO time can be hidden by relatively longer compute time. In contrast, NNs with more IO-bound layers, the IO overhead is higher since the relatively longer IO time cannot be hidden by compute time.
\end{itemize}

\paragraph{Tradeoffs in buffer/tile size, the number of buffers, IO traffic/time, and parallelism}
Given SRAM size, the buffer/tile size and the number of buffers can be decided, and their tradeoffs effects overall IO traffic/time and parallelism in swapping.
\begin{itemize}

	\item 
	\textit{Large buffer/tile size leads to low IO traffic/time, but limits execution parallelism}: Given an NN and SRAM size,
	large buffer/tile size leads to small number of tiles, and hence low IO traffic.
	The overall IO time is short due to less IO traffic, but the execution parallelism is low due to small number of buffers.

	\item 
	\textit{Small buffer/tile size leads to high execution parallelism, but increases overall IO traffic/time}:
	Given an NN and SRAM size, small buffer/tile leads to large number of tiles, and hence high IO traffic.
	The overall IO time is long due to high IO traffic and more fin-grained IO tasks, but the execution parallelism is high due to large number of memory buffers, which allow more tiles to co-exist in memory and be processed in parallel.

\end{itemize}

\paragraph{Experimental insights}
We study how these parameters affect swapping performance on MCU with experiments, and we have the following findings:
\begin{itemize}
	\item \textit{Increasing buffer/tile size can significantly reduce the number of IO tasks and overall IO time.}
	
	The number of IO tasks drops as buffer/tile size increases. Figure~\ref{fig:swapping-tasks} shows the number of IO/compute tasks of NNs under different buffer sizes.
	For instance, when buffer/tile size increases from 16 KB to 128 KB, the number of IO tasks (Grey bars in Figure~\ref{fig:swapping-tasks}) of VGG, AlexNet, and MobileNet drops from 85024 to 2248, from 68040 to 5390, and from 3190 to 870 separately.
	
	Overall IO time drops as buffer/tile size increases. As the IO time (gray bar) shown in Figure~\ref{fig:vgg-512k}, Figure~\ref{fig:alexnet-512k}, and Figure~\ref{fig:mobilenet-512k}, where SRAM size is 512 KB. When buffer/tile size increases from 16 KB to 128 KB, the overall IO time of VGG, AlexNet, and MobileNet drops from 257.824s to 52.4849s, from 27.4765s to 13.6731s, and from 167.183s to 10.6669s separately.
	The same pattern can also be observed when using larger SRAM sizes in Figure~\ref{fig:swapping-latency}.

	\item \textit{Parallel execution can reduce IO overhead, especially when there are larger numbers of buffers.} 
	
	When there are more memory buffers, more IO/compute tasks can be executed in parallel, and hence more IO time can be hidden by compute time.
	For instance, the white and yellow bars in Figure~\ref{fig:vgg-512k} show the sequential execution time and parallel execution time under different buffer sizes (different number of buffers). When buffer/tile size increases from 16 KB to 128 KB, the number of buffers drops from 32 to 4, and IO time reduced by parallel execution drops from 251s to 10s (compared to sequential execution).
	The same pattern can also be observed in other NNs in Figure~\ref{fig:swapping-latency}.

	\item \textit{Given SRAM size, comparing to small buffer/tile size with high parallelism, large buffer/tile size with low parallelism incurs much lower IO overhead in swapping.}

	Both large buffer/tile size (small number of buffers) and high parallelism can reduce IO overhead, but they are in conflict and cannot be achieved at the same time. We observe that the former one can reduce more IO time then the later one.
	
	MobileNet is IO-intensive NN, parallel execution cannot reduce IO overhead much even with more buffers (smaller buffer/tile size, e.g, 16 KB). However, increasing buffer size can reduce IO time from 167s to 16s when buffer size increases from 16KB to 128KB, as shown in Figure~\ref{fig:mobilenet-512k}.
	
	The same pattern also can be observed in AlexNet and VGG, but benefit of choosing large buffer/tile size is not as significant as MobileNet because they are less IO-intensive. 
	For these two NNs, parallel execution plays a bigger role to hide IO time when buffer size is small, while low overall IO tasks/time plays a bigger role when buffer size is large. Overall, large buffer/tile size still overshadows the benefit of parallelism.

\end{itemize}

\subsection{Impact on per-frame delays}

\textit{\textbf{Implication:} With large buffer/tile size, NNs with a small fraction of IO-bound layers see negligible delay increase; NNs with more IO-bound layers see modest delay increase.}

Within a compute-bound layer, MCU can execute IO and computation for consecutive tiles simultaneously (as these tiles are independent), 
completely hiding the IO delay behind the much longer computation delay.
Within an IO-bound layer, IO and compute for consecutive tiles can happen simultaneously as well, but the long IO delay cannot be totally hidden by relatively shorter compute delay.
For other layers, e.g. relu/pooling, the IO/compute delay is insignificant.


As such, the increased delay of an NN due to swapping is mainly determined by the proportion of IO-bound layers' IO delay to all layers' total compute delay. 
The increased delay for NNs with less IO-bound layers is negligible. 
As VGG shown in Table~\ref{tab:cnn-study}, only 2 out of 13 layers are IO-bound, 
leading to only about 6.9\% increased delay as shown in Figure~\ref{fig:vgg-512k} -- Figure ~\ref{fig:vgg-8m} (Yellow vs. Black bars).
The increased delay for NNs with more IO-bound layers is modest. 
As AlexNet and MobileNet show in Table~\ref{tab:cnn-study}, 3 of 5 and 13 of 28 layers are IO-bound, leading to 50\% and 150\% increased delay when buffer/tile size is as large as 128 KB, as shown in Figure~\ref{fig:alexnet-512k} -- Figure ~\ref{fig:mobilenet-8m} (Yellow vs. Black bars).
Overall, the increased delay due to swapping is negligible for compute-intensive NNs and modest for IO-intensive NNs.

\paragraph{\textit{Implication}: }\textit{Insight for hardware designer: increasing SRAM size only increases cost, but cannot improve the latency much in swapping.} 

As shown in Figure~\ref{fig:swapping-latency}, the latency of VGG, AlexNet, and MobileNet does not decrease much as the SRAM size increases. 
For given buffer size, using larger SRAM can increase the number of buffers, and hence can increase parallelism. However increasing SRAM size and the number of buffers cannot help much, because the gap between the number of tasks and the number of buffer is too large (100$\times$ gap). 
For instance, the number of IO tasks in MobileNet is 55877 (Figure~\ref{fig:mobilenet-tasks}) when buffer size is 16 KB, but the number of buffers only increases from 32 to 512 (100$\times$ smaller than 55877) when SRAM size increase from 512KB to 8MB.

\begin{figure*}
    \centering
    \begin{subfigure}[b]{0.23\textwidth}
        \includegraphics[width=\textwidth]{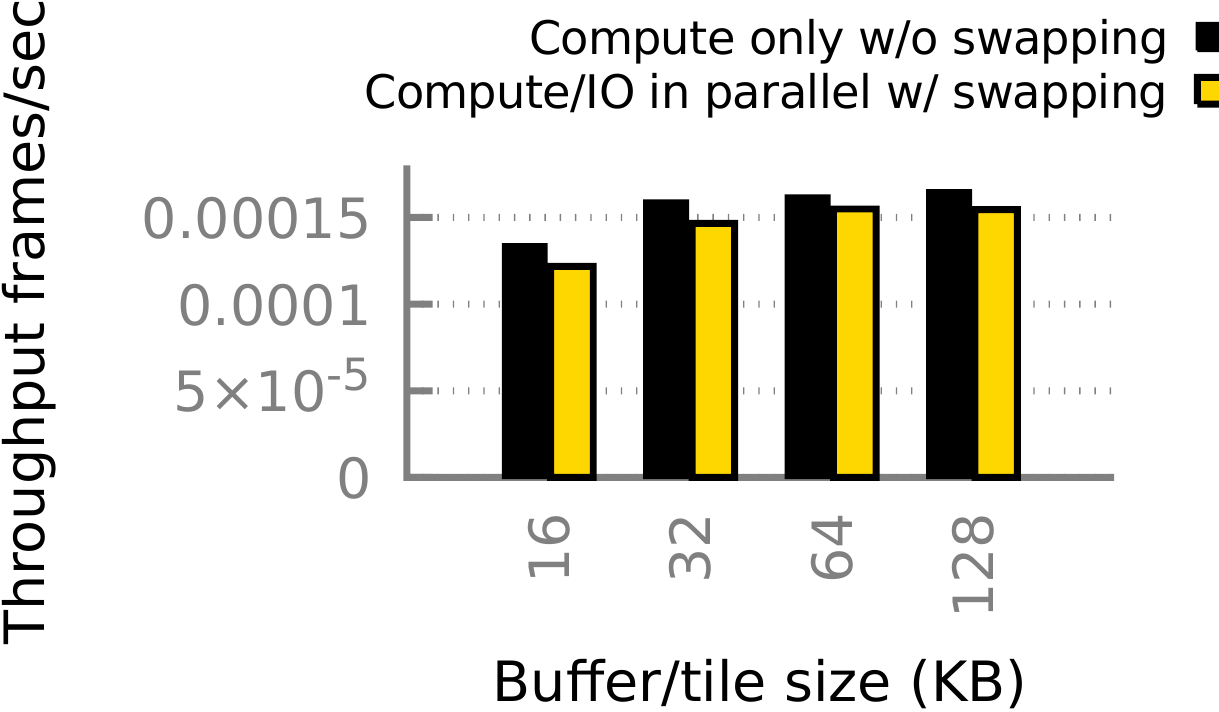}
        \caption{VGG, SRAM 512KB}
        \label{fig:vgg-512k-tput}
    \end{subfigure}
    ~ 
    \begin{subfigure}[b]{0.23\textwidth}
        \includegraphics[width=\textwidth]{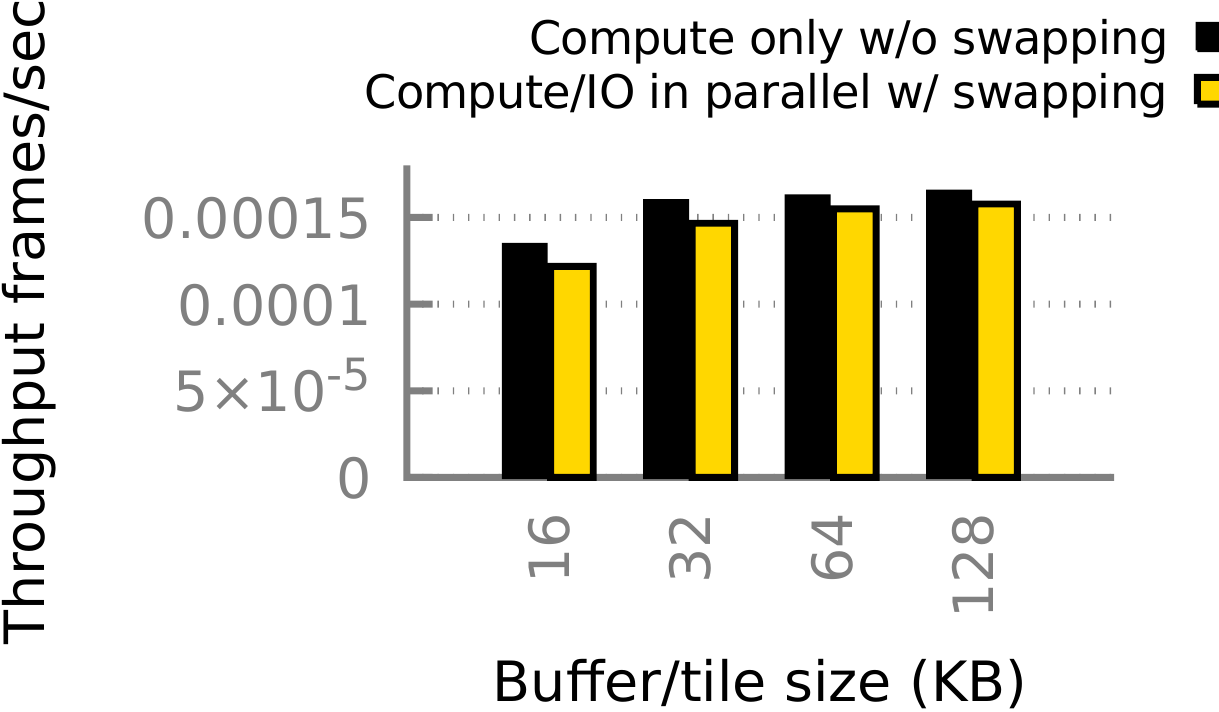}
        \caption{VGG, SRAM 1MB}
        \label{fig:vgg-1m-tput}
    \end{subfigure}
    ~ 
    \begin{subfigure}[b]{0.23\textwidth}
        \includegraphics[width=\textwidth]{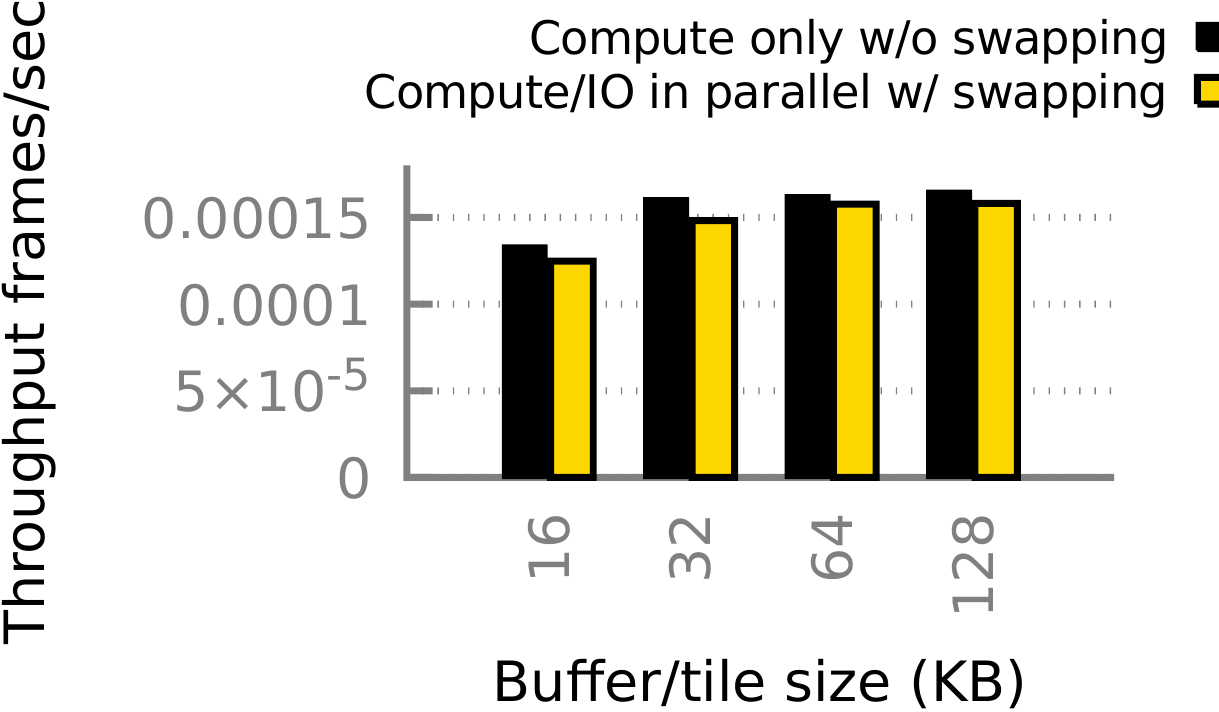}
        \caption{VGG, SRAM 4MB}
        \label{fig:vgg-4m-tput}
    \end{subfigure}
    ~ 
    \begin{subfigure}[b]{0.23\textwidth}
        \includegraphics[width=\textwidth]{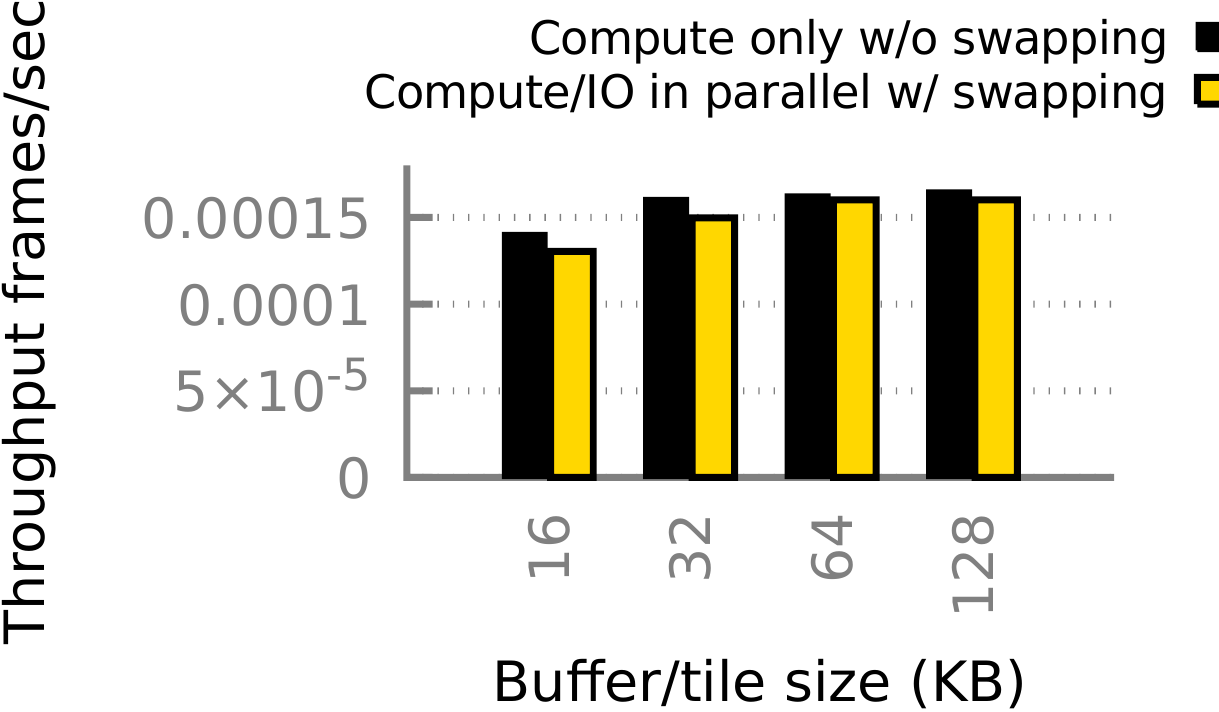}
        \caption{VGG, SRAM 8MB}
        \label{fig:vgg-8m-tput}
    \end{subfigure}    
    
    
    \begin{subfigure}[b]{0.23\textwidth}
        \includegraphics[width=\textwidth]{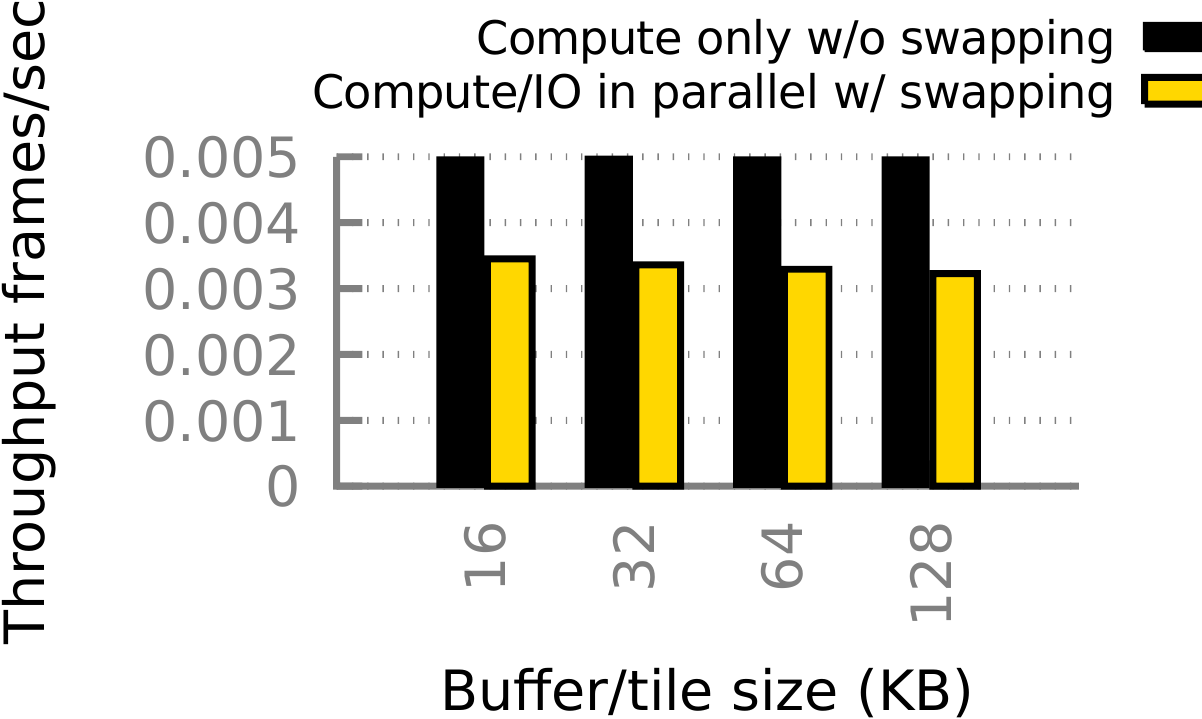}
        \caption{AlexNet, SRAM 512KB}
        \label{fig:alexnet-512k-tput}
    \end{subfigure}
    ~ 
    \begin{subfigure}[b]{0.23\textwidth}
        \includegraphics[width=\textwidth]{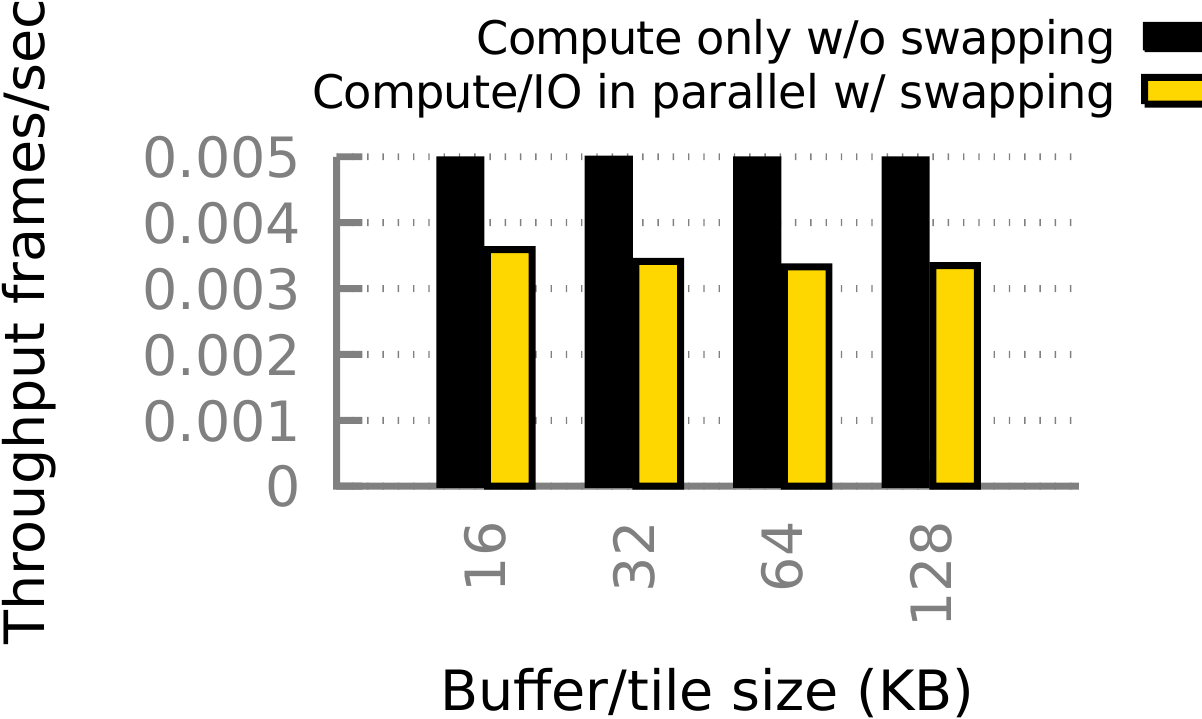}
        \caption{AlexNet, SRAM 1MB}
    \end{subfigure}
    ~ 
    \begin{subfigure}[b]{0.23\textwidth}
        \includegraphics[width=\textwidth]{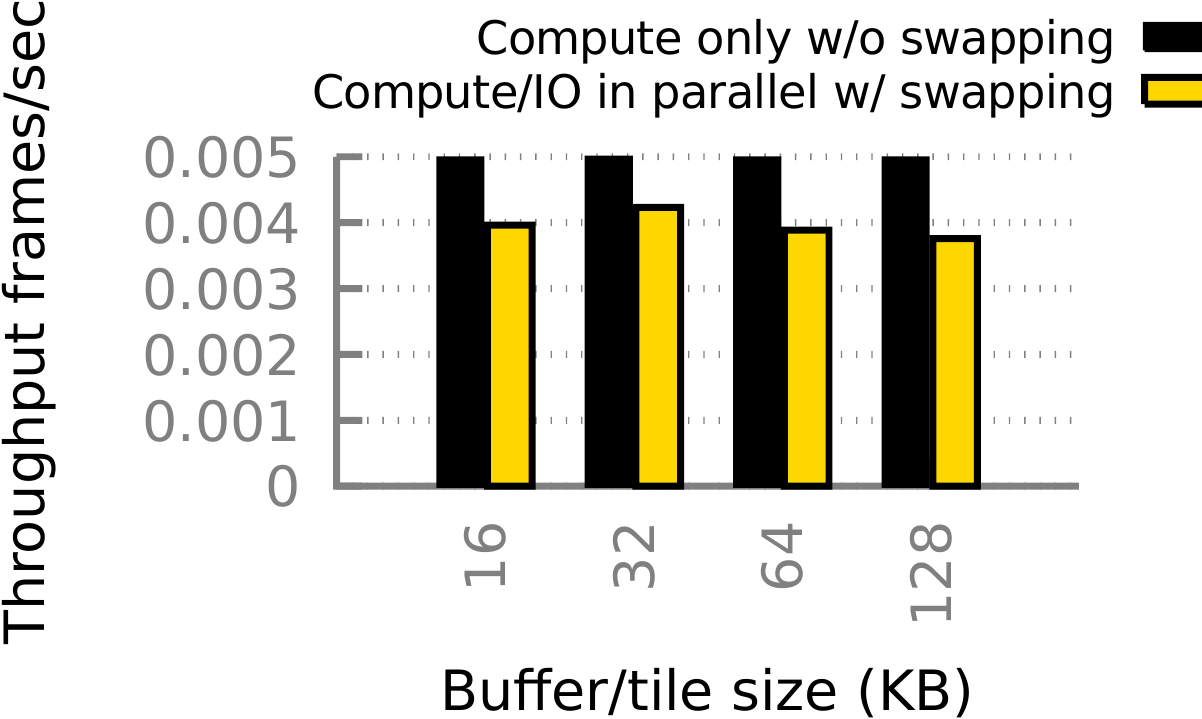}
        \caption{AlexNet, SRAM 4MB}
        \label{fig:alexnet-4m-tput}
    \end{subfigure}
    ~ 
    \begin{subfigure}[b]{0.23\textwidth}
        \includegraphics[width=\textwidth]{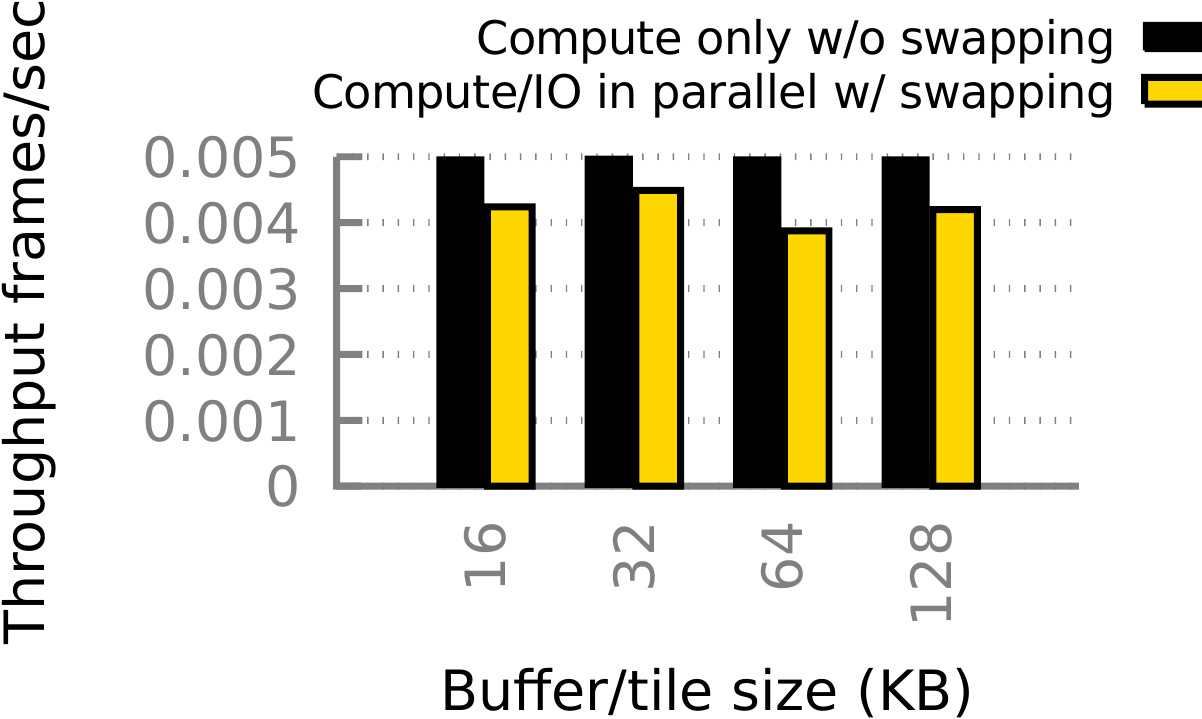}
        \caption{AlexNet, SRAM 8MB}
        \label{fig:alexnet-8m-tput}
    \end{subfigure}
    
    
    \begin{subfigure}[b]{0.23\textwidth}
        \includegraphics[width=\textwidth]{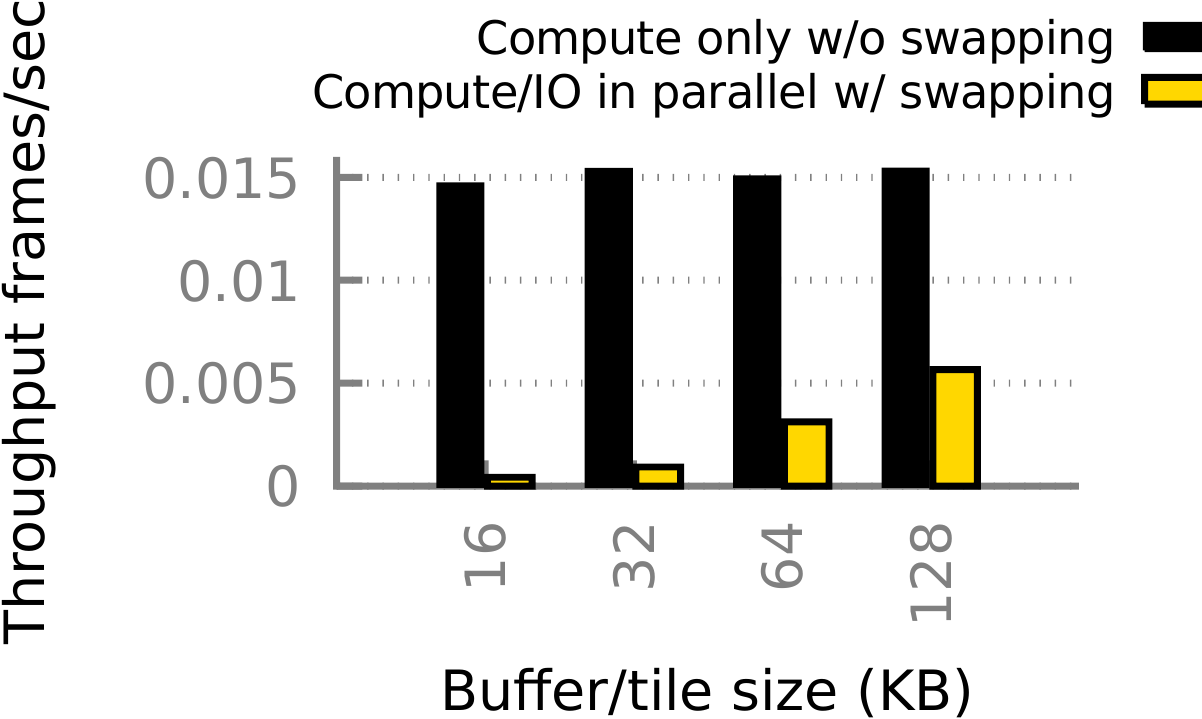}
        \caption{Mobilenet, SRAM 512KB}
        \label{fig:mobilenet-512k-tput}
    \end{subfigure}
    ~ 
    \begin{subfigure}[b]{0.23\textwidth}
        \includegraphics[width=\textwidth]{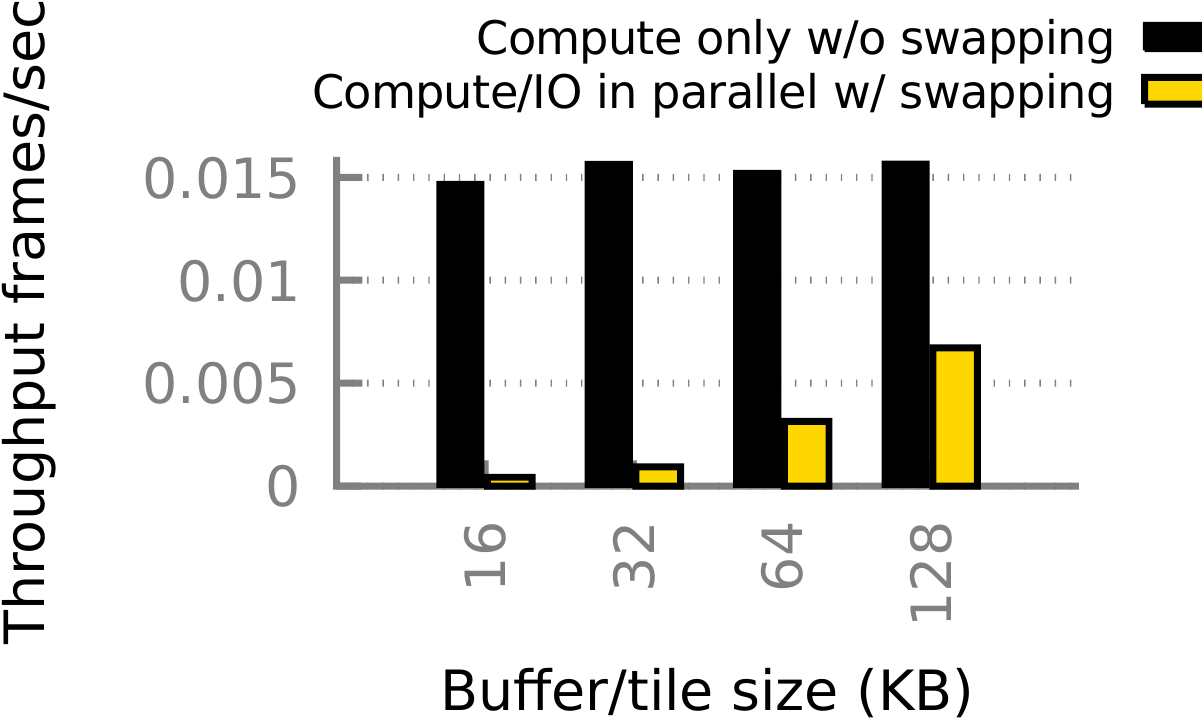}
        \caption{Mobilenet, SRAM 1MB}
        \label{fig:mobilenet-1m-tput}
    \end{subfigure}
    ~ 
    \begin{subfigure}[b]{0.23\textwidth}
        \includegraphics[width=\textwidth]{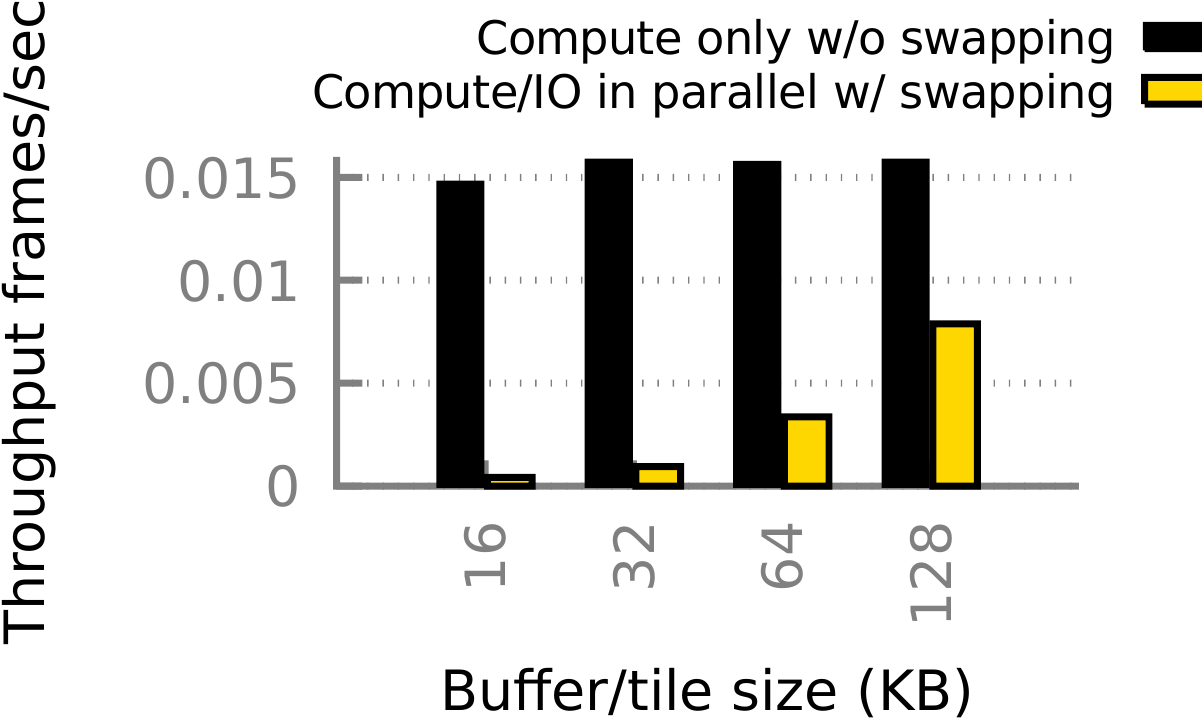}
        \caption{Mobilenet, SRAM 4MB}
        \label{fig:mobilenet-4m-tput}
    \end{subfigure}
    ~ 
    \begin{subfigure}[b]{0.23\textwidth}
        \includegraphics[width=\textwidth]{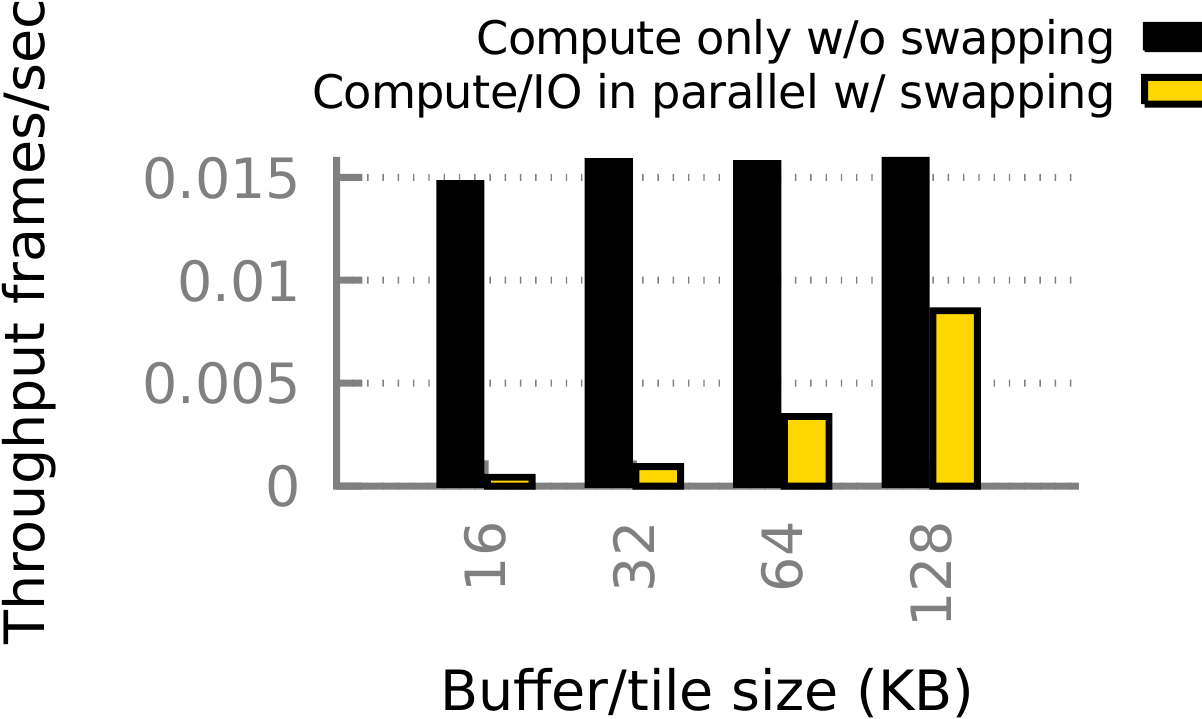}
        \caption{Mobilenet, SRAM 8MB}
        \label{fig:mobilenet-8m-tput}
    \end{subfigure}

    \caption{Swapping throughput of NNs under diffent SRAM sizes and buffer/tile sizes.} 
    \label{fig:swapping-throughput}
\end{figure*}

\subsection{Impact on NN throughput}

\textit{\textbf{Implication:} With large buffer/tile size, NNs see negligible or modest throughput loss.}

NNs with negligible delay increase will also see negligible throughput loss when processing a stream of frames, since the IO time can be hidden by the relatively longer compute time.
For instance, the throughput loss is only 3\% for VGG as shown in Figure~\ref{fig:vgg-8m-tput}, where buffer/tile size is 128 KB and SRAM size is 8 MB.

For those NNs seeing higher delay increase, the throughput loss is relatively higher, since the longer IO time cannot hidden by the relatively shorter compute time.
Although MCU can reduce throughput loss by
exploiting parallelism, but not much due to the limited number of buffers.
For instance, the throughput loss for AlexNet and MobileNet is 15.7\% and 46.4\% as shown in Figure~\ref{fig:alexnet-8m-tput} and Figure~\ref{fig:mobilenet-8m-tput} where buffer size is 128 KB and SRAM size is 8 MB.

\paragraph{\textit{Implication}: }\textit{Cross-frame (pipeline) parallelism cannot improve throughput much due to the limited number of buffers, even if increasing SRAM size.
}

\begin{figure}[t!]
\centering
    \includegraphics[width=0.48\textwidth{}]{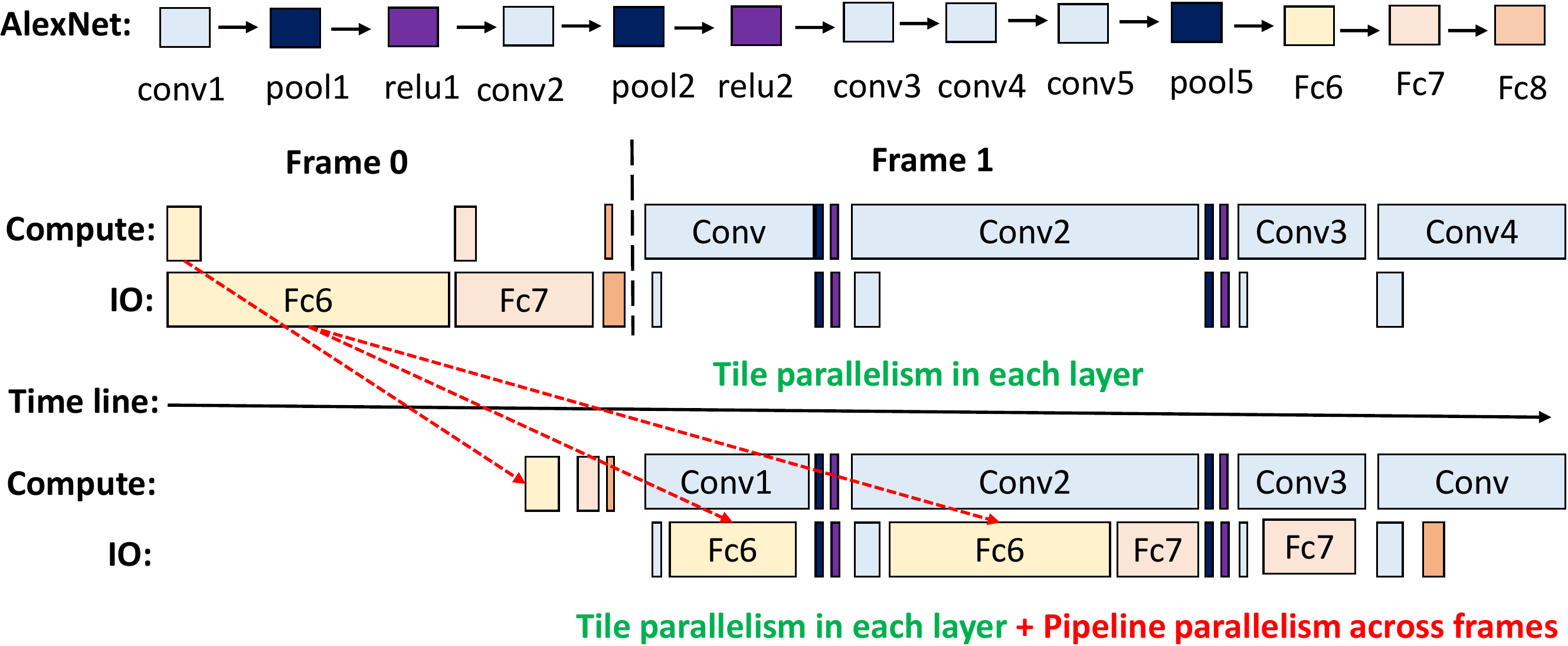} 
    \caption{AlexNet: tile parallelism for low delay and pipeline parallelism for high throughput.}
    \label{fig:alexnet-pipeline-parallelism}
\end{figure}

A common pattern in an NN is that one or more compute-bound layers followed by one or more IO-bound layers, i.e. a \textit{pipeline} with interleaved compute-bound and IO-bound stages. 
For instance, the AlexNet in Figure~\ref{fig:alexnet-128k}, conv1-5 ( compute-bound stage) is followed by fc6-8 (IO-bound stage).
When executing NN on a sequence of frames, MCU can overlap IO/compute-bound stages of adjacent frames, hence hiding the IO delays that cannot be hidden at the layer/tile levels with each frame. 
As shown in Figure~\ref{fig:alexnet-pipeline-parallelism}, MCU can swap for frame 0’s FC layers while computing Frame 1’s Conv layers, leading high MCU/IO utilization and throughput.

However, such parallelism that overlaps IO/compute-bound stages in adjacent frames cannot be fully exploited on MCUs with tiny SRAM due to the limited number of memory buffers.
As Figure~\ref{fig:swapping-throughput} shown, the throughput of VGG, AlexNet, and Mobilenet does not increase much as the SRAM size becomes larger.
Because of the same reason as in latency above, the gap between the number of tasks and the number of buffers is too large (1000$\times$). 
For instance, the number of IO tasks in MobileNet is 558770 (10 frames, 55877 IO tasks in each frame shown in Figure~\ref{fig:mobilenet-tasks}) when buffer size is 16 KB, but the number of buffers only increase from 32 to 512 (1000$\times$ smaller than 558770) when SRAM size increases from 512 KB to 8 MB.
The small number of buffers have been consumed by one frame, so other frames cannot get buffers to be executed in parallel.

\paragraph{\textit{Implication}: }\textit{Increasing buffer/tile size leads to higher throughput than increasing parallelism.}

Same as the tradeoff in latency, given SRAM size: if the buffer/tile size is large (the number of buffer is small), the overall IO time is shot but parallelism is low; if the buffer/tile size is small (the number of buffer is large), the overall IO time is long but the parallelism is high.
Overall, large buffer/tile size leads to higher throughput than small buffer/tile with high parallelism, especially for NNs that have more IO-bound layers.
For instance, MobileNet has more IO-bound layers, and its throughput increase 20$\times$ when buffer size increases from 16 KB to 512 KB (although parallelism drops due to less buffers) as shown in Figure~\ref{fig:mobilenet-8m-tput}.
While VGG and AlexNet have relatively less IO-bound layers, and their throughput does not change much when increasing buffer size, as shown in Figure~\ref{fig:vgg-8m-tput} and Figure~\ref{fig:alexnet-8m-tput}. The reason is that parallelism is high when buffer/tile size is small, and overall IO time is short when buffer size is large.

\subsection{Impact on flash durability}
\label{sec:durability}

\paragraph{\textit{Implication}: }\textit{SD card sees negligible durability loss, and its lifetime could be years or tens of years with swapping.}

The amount of data written to SD card per frame is not large because NN layers are read-most,
and the write frequency is low due to the long execution time on slow MCU.

\paragraph{Modest write rate}
For a given NN and SRAM size, the amount of data written to SD card is determined by the frame rate (reciprocal of delay per frame) and the amount of data to write per frame (upper bound is the sum of output feature maps of all layers), 
which have negative correlations: 
(1) for large NNs, frame rate is low but the amount of data to write per frame is large;
(2) for small NNs, frame rate is high but the amount of data to write per frame is small.
Therefore, no matter an NN is large or small, the data written per day won't be large.
For instance, swapping writes only 2.0/2.8 GB for VGG16/AlexNet per day. 
Even for the extreme case, MobileNet, which has high frame rate and relatively large feature maps to write, swapping writes 123 GB per day.

\paragraph{SD card has long lifetime even with swapping}
SD card is build up of many cells, which have limited write cycles~\cite{kingston}.
As the capacity is becoming larger~\cite{microsdcapacity}, the durability budget is keeping increasing.
The study~\cite{sdtest} 
keeps writing 24/7 as fast as possible to 40 4 GB SD cards, 
and 1, 20, and 40 of 40 cards observe the first failures after writing 6.5 TB, 9 TB, and 12.5 TB of data to them.
Based on their results, the first cell is only expected to fail on a 64 GB SD card after running MobileNet, AlexNet, and VGG16 for 
2.4 -- 4.5, 104 -- 200, and 145 -- 280 years,
and 50\% of cells fail (10K cycles per cell~\cite{flashlife, microsd64g}) only after running for 7.5, 328, and 460 years.
\subsection{Impact on system energy}
\paragraph{\textit{Implication}: }\textit{Swapping adds modest energy consumption to an already busy MCU.} 



We estimate the \textit{worst-case} energy overhead due to swapping. 
Our test platform is an STM32F746NG-Discovery board (ARM Cortex-M7 at 216 MHz; 340 KB SRAM) with an external power meter~\cite{usbtester}. 
We run two benchmarks. 
(1) \textit{in-core} emulates NN executions with an infinite amount of memory: 
it runs NN compute~\cite{cmsisnn} for 1000 iterations. 
(2) \textit{out-of-core} emulates NN executions with the most intensive IO traffic in parallel to the compute: 
it executes the same amount of compute with an IO thread repeatedly flushing data blocks to SD card  
Each data block is 100 KB (close to tile size); 
the flush is asynchronous using the MCU's DMA engine. 
Note that the IO traffic in real applications is less intensive (which will not keep writing all the time) than our benchmarks, so the energy we measure is the \textit{worst-case} energy consumption that is higher than real cases.

Our measurement shows that: the additional IO workloads increases the system energy by 42\%, from 0.07 Wh (in-core) to 0.10 Wh (out-of-core); 
the total execution time goes from 178 sec to 213 sec. 
Our obsevations are: 
(1) The \textit{actual} energy overhead in out-of-core NNs is likely much less:
while the \textit{out-of-core} benchmark keeps IO always busy, the actual out-of-core NNs exercise IO intermittently (\S\ref{sec:io-compute}) because most NN layers are likely compute-bound. 
(2) We attribtute the modest energy overhead to the incremental nature of system energy: when an MCU-based device is already busy executing compute, its most power-hungry hardware -- cores, interconnect, SRAM, and regulators -- is already activated; 
executing IO, which activiates an SD card and the MMC controller in addition, adds to the energy but not much.

\subsection{Out-of-core data security and safety}
Compared to storing NN data in on-chip SRAM, (temporarily) storing it off-chip is more vulnerable to physical attacks~\cite{hackingsd}: 
adversaries may learn or corrupt the data by tapping into the IO bus between MCU and the SD card, or the SD card itself. 
Fortunately, by encrypting NN data before swapping out, MCU can ensure the data to be confidential and integral; the overhead is linear to the data amount. 
Hardware crypto, such as for ASE~\cite{armcrypto,schwabe2016all}, is already common on modern MCUs. 
Its computation overhead is comparable to (or even less than) the least intensive NN compute (e.g. FC layers). 


Compared to SRAM, SD cards are less durable. 
Yet, it is known that a SD card rarely fails as a whole but seeing a gradual increase number of corrupted cells over time~\cite{reliablesd}. 
Cell corruption is often silent, i.e. a read value simply differs from what was written last time. 
Fortunately, MCU can detects such failures with hash-based integrity checking. 
With specialized hardware on MCUs, computing hash is no more expensive than the least intensive NN compute~\cite{armcrypto}. 
Upon detection of bad cells, the MCU can recompute the most recent NN layer and recover the corrupted out-of-core data. 
 \section{Related work}

\paragraph{Implications on model compression}
Existing work on tinyML tries to run NNs on MCUs by reducing memory footprint, such as model compression~\cite{prune, deepcompression, prunefilter}, parameter quantization~\cite{limitedprecision}, designing tiny NNs from scratch~\cite{bmxnet}, as well as automation of these procedures~\cite{lin2020mcunet}. 
However, they give away model accuracy or generality at varying degrees. 
In order for an NN to fit into the MCU memory,  the NN either becomes substantially inaccurate or too specialized.
In contrast, our swapping solution doesn't incur accuracy and generality loss.
Our solution boosts design freedom in tinyML, 
where memory limit was considered as the primary motivation for model compression. 
With the removal of such a limit, developers now have the choice of run large NNs without compression, retaining full model accuracy. 
Even in case of model compression is warranted, e.g. for faster NN execution, 
developers now have a wider selection of \textit{baseline} NNs, including the ones with orders of higher memory footprints than MCUs.

\paragraph{Relation to prior swapping systems}
Prior work enables out-of-core NN training with large batches on GPU/CPU memory systems~\cite{hayakawa2020out, swapadvisor, vdnn, dynamic, splitcnn}, 
but they cannot address the unique challenge on MCU that even a single layer exceeds main memory during NN inference.
Prior work, e.g., Scratch-Pad~\cite{dominguez2005heap} , proposes generic technique to swap data between SRAM and DRAM (not SD) for embedded devices. However, they don't leverage NN characteristics to optimize swapping, and 
they don't answer how swapping affects SD card lifetime, execution slowdown, energy consumption, and data security for NN applications.
This paper presents the first study on these questions and shows that swapping is feasible without much overhead.

\paragraph{Complement to existing inference framework}
Tensorflow Lite Micro~\cite{tflitemicro} is a framework for running NN inference on embedded devices. 
CMSIS-NN~\cite{cmsisnn} provides optimized NN kernels for ARM Cortex-M MCUs.
SONIC~\cite{intelligenceedge} supports intermittent computing for NN inference on MUCs.
TVM~\cite{tvm} can generate optimized code for NNs on MCUs.
However, none of them supports NNs whose memory footprints are larger than physical memory on MCUs. 
Our out-of-core solution is a complement to existing frameworks. 
It can be used in conjunction with them and expand their design space.

\section{Conclusions}

This paper advocates enabling large NNs on tiny MCUs without losing accuracy by swapping data to SD card.
With the parallel scheduler that overlaps IO and compute tasks to hide IO overhead, our study shows that none of SD card durability loss, execution slowdown, energy consumption, or data security is an issue. 
We find that an MCU with hundreds of KBs SRAM can execute NNs with a few hundreds MBs of memory footprint (a 1000$\times$ gap).
Out-of-core execution expands the scope of NN applications on MCUs.

\bibliographystyle{ACM-Reference-Format}
\bibliography{bib/video,bib/edge-nn,bib/sec,bib/failure}

\begin{thebibliography}{10}

\bibitem{flashlife}
Every thing you need to know about slc, mlc, and tlc nand flash.
\newblock
  \url{https://www.mydigitaldiscount.com/everything-you-need-to-know-about-slc-mlc-and-tlc-nand-flash.html}.

\bibitem{microsdcapacity}
History and evolution of memory cards.
\newblock
  \url{https://koofr.eu/blog/posts/history-and-evolution-of-memory-cards}.

\bibitem{kingston}
Kingston flash memory guide.
\newblock
  \url{https://media.kingston.com/pdfs/MKF_283.1_Flash_Memory_Guide_EN.pdf}.

\bibitem{microsdbw}
microsd card benchmarks.
\newblock \url{https://www.pidramble.com/wiki/benchmarks/microsd-cards}.

\bibitem{nxpmcu}
Nxp general purpose microcontrollers.
\newblock
  \url{https://www.nxp.com/products/processors-and-microcontrollers/arm-microcontrollers/general-purpose-mcus:GENERAL-PURPOSE-MCUS}.

\bibitem{sramiot}
The role of srams in nextgen iot and wearable embedded designs.
\newblock
  \url{https://www.embedded.com/the-role-of-srams-in-nextgen-iot-and-wearable-embedded-designs/}.

\bibitem{roofline}
Roofline model.
\newblock \url{https://en.wikipedia.org/wiki/Roofline_model}.

\bibitem{sdtest}
Sd cart testing.
\newblock
  \url{https://support.embeddedarm.com/support/solutions/articles/22000202866-sd-card-testing}.

\bibitem{stmcu}
Stm32 32-bit arm cortex mcu.
\newblock
  \url{https://www.st.com/en/microcontrollers-microprocessors/stm32-32-bit-arm-cortex-mcus.html}.

\bibitem{reliablesd}
Reliable sd-based block storage.
\newblock
  \url{https://support.embeddedarm.com/support/solutions/articles/22000202867-reliable-sd-based-block-storage},
  2017.

\bibitem{cortexm}
Arm cortex-m.
\newblock \url{https://en.wikipedia.org/wiki/ARM_Cortex-M}, 2020.

\bibitem{hackingsd}
The exploration and exploitation of an sd memory card.
\newblock \url{http://bunniefoo.com/bunnie/sdcard-30c3-pub.pdf}, 2020.

\bibitem{flops}
Floating point operations per second.
\newblock \url{https://en.wikipedia.org/wiki/FLOPS}, 2020.

\bibitem{introtinyml}
An introduction to tinyml.
\newblock
  \url{https://towardsdatascience.com/an-introduction-to-tinyml-4617f314aa79},
  2020.

\bibitem{nuruai}
Nuru ai expansion: Supporting farmers to diagnose crop diseases.
\newblock
  \url{https://blog.plantwise.org/2020/03/13/nuru-ai-expansion-supporting-farmers-to-diagnose-crop-diseases/},
  2020.

\bibitem{armcrypto}
Performance of state-of-the-art cryptography on arm-based microprocessors.
\newblock
  \url{https://csrc.nist.gov/csrc/media/events/lightweight-cryptography-workshop-2015/documents/presentations/session7-vincent.pdf},
  2020.

\bibitem{stm32}
Stmicroelectronics stm32 family.
\newblock \url{https://en.wikipedia.org/wiki/STM32}, 2020.

\bibitem{microsd64g}
Transcend industrial temp microsd 64 gb.
\newblock
  \url{https://cdn.transcend-info.com/products/images/modelpic/574/EN_USDC10I_PS_2020.pdf},
  2020.

\bibitem{usbtester}
Usb c power meter tester.
\newblock
  \url{https://www.amazon.com/gp/product/B07X3HST7V/ref=ppx_yo_dt_b_asin_title_o00_s00?ie=UTF8&psc=1},
  2020.

\bibitem{convnet-burden}
Samuel albanie.
\newblock Estimates of memory consumption and flop counts for various
  convolutional neural networks.
\newblock \url{https://github.com/albanie/convnet-burden}, 2021.

\bibitem{layerfusion}
Manoj Alwani, Han Chen, Michael Ferdman, and Peter Milder.
\newblock Fused-layer cnn accelerators.
\newblock In {\em The 49th Annual IEEE/ACM International Symposium on
  Microarchitecture}, page~22. IEEE Press, 2016.

\bibitem{tvm}
Tianqi Chen, Thierry Moreau, Ziheng Jiang, Lianmin Zheng, Eddie Yan, Haichen
  Shen, Meghan Cowan, Leyuan Wang, Yuwei Hu, Luis Ceze, et~al.
\newblock Tvm: An automated end-to-end optimizing compiler for deep learning.
\newblock In {\em 13th USENIX Symposium on Operating Systems Design and
  Implementation (OSDI 18)}, pages 578--594, 2018.

\bibitem{tflitemicro}
Robert David, Jared Duke, Advait Jain, Vijay~Janapa Reddi, Nat Jeffries, Jian
  Li, Nick Kreeger, Ian Nappier, Meghna Natraj, Shlomi Regev, et~al.
\newblock Tensorflow lite micro: Embedded machine learning on tinyml systems.
\newblock {\em arXiv preprint arXiv:2010.08678}, 2020.

\bibitem{statofnnpruning}
Jonathan Frankle John~Guttag Davis~Blalock, Jose Javier Gonzalez~Ortiz.
\newblock What is the state of neural network pruning?
\newblock In {\em MLSys}, 2020.

\bibitem{dominguez2005heap}
Angel Dominguez, Sumesh Udayakumaran, and Rajeev Barua.
\newblock Heap data allocation to scratch-pad memory in embedded systems.
\newblock {\em Journal of Embedded Computing}, 1(4):521--540, 2005.

\bibitem{intelligenceedge}
Graham Gobieski, Brandon Lucia, and Nathan Beckmann.
\newblock Intelligence beyond the edge: Inference on intermittent embedded
  systems.
\newblock In {\em Proceedings of the Twenty-Fourth International Conference on
  Architectural Support for Programming Languages and Operating Systems}, pages
  199--213. ACM, 2019.

\bibitem{limitedprecision}
Suyog Gupta, Ankur Agrawal, Kailash Gopalakrishnan, and Pritish Narayanan.
\newblock Deep learning with limited numerical precision.
\newblock In {\em International Conference on Machine Learning}, pages
  1737--1746, 2015.

\bibitem{deepcompression}
Song Han, Huizi Mao, and William~J Dally.
\newblock Deep compression: Compressing deep neural networks with pruning,
  trained quantization and huffman coding.
\newblock {\em arXiv preprint arXiv:1510.00149}, 2015.

\bibitem{hayakawa2020out}
Akio Hayakawa and Takuya Narihira.
\newblock Out-of-core training for extremely large-scale neural networks with
  adaptive window-based scheduling.
\newblock {\em arXiv preprint arXiv:2010.14109}, 2020.

\bibitem{resnet}
Kaiming He, Xiangyu Zhang, Shaoqing Ren, and Jian Sun.
\newblock Deep residual learning for image recognition.
\newblock In {\em Proceedings of the IEEE conference on computer vision and
  pattern recognition}, pages 770--778, 2016.

\bibitem{mobilenets}
Andrew~G Howard, Menglong Zhu, Bo~Chen, Dmitry Kalenichenko, Weijun Wang,
  Tobias Weyand, Marco Andreetto, and Hartwig Adam.
\newblock Mobilenets: Efficient convolutional neural networks for mobile vision
  applications.
\newblock {\em arXiv preprint arXiv:1704.04861}, 2017.

\bibitem{swapadvisor}
Chien-Chin Huang, Gu~Jin, and Jinyang Li.
\newblock Swapadvisor: Pushing deep learning beyond the gpu memory limit via
  smart swapping.
\newblock In {\em Proceedings of the Twenty-Fifth International Conference on
  Architectural Support for Programming Languages and Operating Systems}, pages
  1341--1355, 2020.

\bibitem{splitcnn}
Tian Jin and Seokin Hong.
\newblock Split-cnn: Splitting window-based operations in convolutional neural
  networks for memory system optimization.
\newblock In {\em Proceedings of the Twenty-Fourth International Conference on
  Architectural Support for Programming Languages and Operating Systems}, pages
  835--847, 2019.

\bibitem{alexnet}
Alex Krizhevsky, Ilya Sutskever, and Geoffrey~E Hinton.
\newblock Imagenet classification with deep convolutional neural networks.
\newblock In {\em Advances in neural information processing systems}, pages
  1097--1105, 2012.

\bibitem{cmsisnn}
Liangzhen Lai, Naveen Suda, and Vikas Chandra.
\newblock Cmsis-nn: Efficient neural network kernels for arm cortex-m cpus.
\newblock {\em arXiv preprint arXiv:1801.06601}, 2018.

\bibitem{prunefilter}
Hao Li, Asim Kadav, Igor Durdanovic, Hanan Samet, and Hans~Peter Graf.
\newblock Pruning filters for efficient convnets.
\newblock {\em arXiv preprint arXiv:1608.08710}, 2016.

\bibitem{lin2020mcunet}
Ji~Lin, Wei-Ming Chen, Yujun Lin, John Cohn, Chuang Gan, and Song Han.
\newblock Mcunet: Tiny deep learning on iot devices.
\newblock {\em arXiv preprint arXiv:2007.10319}, 2020.

\bibitem{vdnn}
Minsoo Rhu, Natalia Gimelshein, Jason Clemons, Arslan Zulfiqar, and Stephen~W
  Keckler.
\newblock vdnn: Virtualized deep neural networks for scalable, memory-efficient
  neural network design.
\newblock In {\em 2016 49th Annual IEEE/ACM International Symposium on
  Microarchitecture (MICRO)}, pages 1--13. IEEE, 2016.

\bibitem{schwabe2016all}
Peter Schwabe and Ko~Stoffelen.
\newblock All the aes you need on cortex-m3 and m4.
\newblock In {\em International Conference on Selected Areas in Cryptography},
  pages 180--194. Springer, 2016.

\bibitem{shen2017fast}
Haichen Shen, Seungyeop Han, Matthai Philipose, and Arvind Krishnamurthy.
\newblock Fast video classification via adaptive cascading of deep models.
\newblock In {\em Proceedings of the IEEE conference on computer vision and
  pattern recognition}, pages 3646--3654, 2017.

\bibitem{vgg}
Karen Simonyan and Andrew Zisserman.
\newblock Very deep convolutional networks for large-scale image recognition.
\newblock {\em arXiv preprint arXiv:1409.1556}, 2014.

\bibitem{siu2018memory}
Kevin Siu, Dylan~Malone Stuart, Mostafa Mahmoud, and Andreas Moshovos.
\newblock Memory requirements for convolutional neural network hardware
  accelerators.
\newblock In {\em 2018 IEEE International Symposium on Workload
  Characterization (IISWC)}, pages 111--121. IEEE, 2018.

\bibitem{googlenet}
Christian Szegedy, Wei Liu, Yangqing Jia, Pierre Sermanet, Scott Reed, Dragomir
  Anguelov, Dumitru Erhan, Vincent Vanhoucke, and Andrew Rabinovich.
\newblock Going deeper with convolutions.
\newblock In {\em Proceedings of the IEEE conference on computer vision and
  pattern recognition}, pages 1--9, 2015.

\bibitem{szegedy2015going}
Christian Szegedy, Wei Liu, Yangqing Jia, Pierre Sermanet, Scott Reed, Dragomir
  Anguelov, Dumitru Erhan, Vincent Vanhoucke, and Andrew Rabinovich.
\newblock Going deeper with convolutions.
\newblock In {\em Proceedings of the IEEE conference on computer vision and
  pattern recognition}, pages 1--9, 2015.

\bibitem{deeptype}
Mengwei Xu, Feng Qian, Qiaozhu Mei, Kang Huang, and Xuanzhe Liu.
\newblock Deeptype: On-device deep learning for input personalization service
  with minimal privacy concern.
\newblock {\em Proceedings of the ACM on Interactive, Mobile, Wearable and
  Ubiquitous Technologies}, 2(4):1--26, 2018.

\bibitem{xu2020approximate}
Mengwei Xu, Xiwen Zhang, Yunxin Liu, Gang Huang, Xuanzhe Liu, and Felix~Xiaozhu
  Lin.
\newblock Approximate query service on autonomous iot cameras.
\newblock In {\em Proceedings of the 18th International Conference on Mobile
  Systems, Applications, and Services}, pages 191--205, 2020.

\bibitem{bmxnet}
Haojin Yang, Martin Fritzsche, Christian Bartz, and Christoph Meinel.
\newblock Bmxnet: An open-source binary neural network implementation based on
  mxnet.
\newblock In {\em Proceedings of the 25th ACM international conference on
  Multimedia}, pages 1209--1212. ACM, 2017.

\bibitem{dynamic}
Yuan Yu, Mart{\'\i}n Abadi, Paul Barham, Eugene Brevdo, Mike Burrows, Andy
  Davis, Jeff Dean, Sanjay Ghemawat, Tim Harley, Peter Hawkins, et~al.
\newblock Dynamic control flow in large-scale machine learning.
\newblock In {\em Proceedings of the Thirteenth EuroSys Conference}, pages
  1--15, 2018.

\bibitem{zhang2017hello}
Yundong Zhang, Naveen Suda, Liangzhen Lai, and Vikas Chandra.
\newblock Hello edge: Keyword spotting on microcontrollers.
\newblock {\em arXiv preprint arXiv:1711.07128}, 2017.

\bibitem{prune}
Michael Zhu and Suyog Gupta.
\newblock To prune, or not to prune: exploring the efficacy of pruning for
  model compression.
\newblock {\em arXiv preprint arXiv:1710.01878}, 2017.

\end{thebibliography}
\end{document}